\documentclass[preprint,aps,tightenlines,floatfix]{revtex4}
\usepackage{epsfig}\usepackage{graphics} \usepackage{amsmath}

\begin{document}

\preprint{
\vbox{
% \hbox{June 2003}
\hbox{ADP-03-114/T552}
\hbox{JLAB-THY-03-30}
\hbox{KSUCNR-203-06}
}}

\vspace*{1cm}

\title{Deep Inelastic Scattering from $A=3$ Nuclei      \\
        and the Neutron Structure Function              \\}

\author{I.R.~Afnan$^1$, F.~Bissey$^2$, J.~Gomez$^3$,
        A.T.~Katramatou$^4$, S.~Liuti$^5$, W.~Melnitchouk$^3$,
        G.G.~Petratos$^4$, A.W.~Thomas$^6$}

\affiliation{$^1$ School of Physical Sciences,
        Flinders University of South Australia,
        Bedford Park, 5042, Australia}
\affiliation{$^2$ Universit\'e de Li\`ege,
        D\'epartement de Physique,
        Institut de Physique B.5,
        Sart Tilman,
        B-4000 Liege 1, Belgium}
\affiliation{$^3$ Jefferson Lab,
        12000 Jefferson Avenue,
        Newport News, VA 23606, USA}
\affiliation{$^4$ Kent State University,
        Kent, OH 44242, USA}
\affiliation{$^5$ Department of Physics,
        University of Virginia,
        Charlottesville, VA 22904, USA,         \\
        and I.N.F.N., Sezione di Roma Tre,
        Via della Vasca Navale,
        00184 Roma, Italy,}
\affiliation{$^6$ Special Research Centre for the
        Subatomic Structure of Matter,
        and Department of Physics and Mathematical Physics,
        University of Adelaide, 5005, Australia \\ \\}

\begin{abstract}
We present a comprehensive analysis of deep inelastic scattering from
$^3$He and $^3$H, focusing in particular on the extraction of the
free neutron structure function, $F_2^n$.
Nuclear corrections are shown to cancel to within 1--2\% for the
isospin-weighted ratio of $^3$He to $^3$H structure functions,
which leads to more than an order of magnitude improvement in the
current uncertainty on the neutron to proton ratio $F_2^n/F_2^p$
at large $x$.
Theoretical uncertainties originating from the nuclear wave function,
including possible non-nucleonic components, are evaluated.
Measurement of the $^3$He and $^3$H structure functions will, in addition,
determine the magnitude of the EMC effect in all $A \leq 3$ nuclei.
\end{abstract}

\maketitle

%%%%%%%%%%%%%%%%%%%%%%%%%%%%%%%%%%%%%%%%%%%%%%%%%%%%%%%%%%%%%%%%%%%%%%%%%
\section{Introduction}

It is a somewhat anomalous situation whereby the nuclear effects in
deep inelastic scattering (DIS) from few-nucleon systems for which the
theoretical descriptions are most easily tractable, namely the deuteron,
helium-3 and tritium, are the least well known experimentally.
For example, the nuclear EMC effect has been extensively studied for
$4 < A \alt 200$ \cite{GST}, but 20 years after the original EMC
observation \cite{EMC} of nucleon structure function modification
in medium, it is still not known for $A=2$ or 3 systems.

The lack of knowledge of the EMC effect in $A < 4$ nuclei has been a
major obstacle to a complete description of the nucleon structure
functions themselves.
The distribution of valence $u$ and $d$ quarks in the proton can be
determined from any two observables containing linear combinations of
$u$ and $d$ quarks, which are usually taken to be the proton and
neutron structure functions, $F_2^p$ and $F_2^n$.
While the proton structure function is quite well constrained for
light-cone momentum fractions $x = Q^2/2M\nu \alt 0.8$, the neutron
$F_2^n$ is usually extracted from data on deuterium, however, beyond
$x \sim 0.5$ the large nuclear corrections can result in uncertainties
of up to $\sim 50\%$ in $F_2^n/F_2^p$ \cite{WHITLOW,FS88,LG,MT}.
Here $Q^2$ and $\nu$ are (minus) the photon virtuality and energy,
and $M$ is the nucleon mass.
Inclusive proton and deuteron data, which have been almost exclusively
been used to constrain the $d/u$ ratio, are therefore unreliable for
determining the neutron structure function beyond $x \sim 0.5$, and
other methods must be sought.

Several alternatives for obtaining an independent linear combination
of $u$ and $d$ quark distributions have been discussed recently, which
could minimize or avoid the problem of nuclear corrections.
These include flavor tagging in semi-inclusive scattering from
hydrogen, in which $\pi^\pm$ production at large $z$ selects $u$ and
$d$ quarks, respectively \cite{SEMIPI}, and parity-violating $\vec e p$
scattering, for which the left--right polarization asymmetry arising
from the $\gamma^*$--$Z$ interference is, at leading order, proportional
to $d/u$ \cite{PARITY}.
Other proposals have utilized the weak charged current to couple
preferentially either $u$ or $d$ flavors, for example asymmetries in
$W$-boson production in $pp$ and $p\bar p$ collisions \cite{W} at
Fermilab or RHIC, or charged current $e^+ p$ deep inelastic scattering
at HERA \cite{HERACC}.
One of the more promising techniques appears to be semi-inclusive DIS
from a deuterium target, with coincidence detection of a low momentum
spectator proton in the target fragmentation region, which maximizes the
likelihood of scattering from a nearly on-shell neutron \cite{BONUS,SPEC}.

In this paper we focus on a novel idea which would not be subject to
the low rates associated with weak current reactions, nor rely on the
validity of factorization of target and current hadrons in the final
state in semi-inclusive scattering.
It involves maximally exploiting the mirror symmetry of $A=3$ nuclei
to extract the $F_2^n/F_2^p$ ratio from the ratio of $^3$He/$^3$H
$F_2$ structure functions \cite{WKSHP}.
Differences in the relative size of nuclear effects in $^3$He and $^3$H
are quite small --- essentially on the scale of charge symmetry breaking
in nuclei --- even though the absolute size of the EMC effect in an $A=3$
nucleus can be relatively large.
Preliminary results for the expected errors in the extraction have been
presented in Ref.~\cite{AFNAN}.
(See also Ref.~\cite{SPINWKSHP}.)
Here we discuss in greater detail the possible theoretical
uncertainties associated with nuclear effects in three-body nuclei,
and experimental considerations relevant for a clean measurement of the
$^3$He/$^3$H structure function ratio.
Some of the latter have been summarized in Ref.~\cite{12GEV}.
In particular, we consider effects such as different nuclear wave
functions, charge symmetry breaking, finite $Q^2$ corrections,
as well as non-nucleonic degrees of freedom such as six-quark clusters,
and explicit nucleon off-shell effects.

In Section~II we motivate the need for new measurements of the free
neutron structure function in the hitherto unexplored kinematic region
at large $x$, and outline the extraction of $F_2^n$ from the
$F_2^{^3{\rm He}}$ and $F_2^{^3{\rm H}}$ structure functions.
A detailed discussion of the theoretical framework and the nuclear
spectral functions is presented in Section~III.
As well as allowing for a relatively clean extraction of the
$F_2^n/F_2^p$ ratio, deep inelastic scattering from $^3$He/$^3$H can
also provide the first indications of the absolute size of the EMC
effect in $A=3$ nuclei.
With the exception of the recent HERMES data \cite{HERMES} at lower $x$
and $Q^2$ on the ratio of $^3$He to $p$ and $d$ cross sections, all
existing data on the nuclear EMC effect are for $A \ge 4$.
Predictions for the EMC ratios in $^3$He and $^3$H based on the
conventional nuclear descriptions are discussed in Section~III.

The sensitivity of the extracted $F_2^n$ to nuclear effects is dealt
with in detail in Section~IV, where in addition to conventional nuclear
models of the $A=3$ system in terms of well-known three-body wave
functions, we examine more speculative models, including those involving
explicit non-nucleon degrees of freedom, in order to assess the possible
model-dependence of the extraction.
We find that for all models which are known to be consistent with
standard nuclear phenomenology, the nuclear effects in the ratio of the
EMC effects in $^3{\rm He}$ and $^3{\rm H}$ cancel to within 1--2\% for
$x \alt 0.8$.
In Section~V we calculate the expected rates at which the $^3$He and
$^3$H cross sections can be determined experimentally at future
facilities, such as Jefferson Lab with 12~GeV electron energy.
Finally, we summarize our findings in Section~VI.

%%%%%%%%%%%%%%%%%%%%%%%%%%%%%%%%%%%%%%%%%%%%%%%%%%%%%%%%%%%%%%%%%%%%%%%%%
\section{Neutron Structure Function and the $A=3$ System}

In this Section we outline the theoretical motivation for determining
precisely the neutron structure function at large $x$, and describe
in detail the method proposed to extract $F_2^n$ from deep inelastic
$^3$He and $^3$H structure functions.

% .......................................................................
\subsection{Neutron Structure and Spin-Flavor Symmetry Breaking}

An accurate determination of the neutron structure function $F_2^n$
is essential for pinning down the momentum dependence of both the
$u$ and $d$ quarks in the nucleon.
While the $u$ quark distribution in the proton is relatively well
determined by the proton $F_2$ data, the $d/u$ ratio at large $x$ is,
at leading order, usually extracted from a ratio of the neutron to
proton structure functions:
\begin{eqnarray}
\label{F2np}
{ F_2^n \over F_2^p }
&=& { 1 + 4 d/u  \over 4 + d/u }\ .
\end{eqnarray}
According to SU(6) symmetry one would expect that $u = 2 d$ for all $x$,
so that $F_2^n/F_2^p = 2/3$, although the data have for a long time been  
known to deviate strongly from this naive expectation beyond $x \sim 0.4$.
A number of different non-perturbative mechanisms have been suggested
\cite{FEYNMAN,CLOSE,CT,CARLITZ,ISGUR,FJ,DUAL,NOSU6} which break SU(6)
symmetry, and most have been able to fit the data in the region of $x$
where $n/p$ can be reliably extracted.

On the other hand, the $x \to 1$ behavior of $F_2^n/F_2^p$ predicted
by the various models depends rather strongly on the assumed dynamics
responsible for the symmetry breaking.
In particular, whether the suppression of the $d$ quark at large $x$
is due to suppression of helicity anti-aligned quarks in the proton,
or non-perturbative interactions which raise the energy of the
scalar-isoscalar diquark components of the proton wave function, the
$x \to 1$ limit of $F_2^n/F_2^p$ can vary from 1/4 \cite{FEYNMAN,CLOSE}
up to 3/7 \cite{FJ}.
Theoretical uncertainties in the currently extracted $F_2^n$ at
large $x$ are comparable to the differences between the $x \to 1$
behaviors.
In particular, whether one corrects for Fermi motion and binding
in the deuteron \cite{MT}, or Fermi motion alone \cite{BR,bodek}, the
extracted $F_2^n$ can appear to approach either of the predicted
limits, as shown in Fig.~1.
(This is reminiscent of the large deuteron wave function model   
dependence of the extracted neutron electric form factor
\cite{PLATCHKOV}.)

Apart from testing non-perturbative QCD dynamics, a very practical
reason for determining large-$x$ distributions is the need to precisely
constrain the input distributions for calculations of cross sections at
high energy colliders.
Uncertainties in parton distributions at large $x$ and modest $Q^2$
translate via perturbative QCD evolution into uncertainties at high
$Q^2$ at lower $x$.
This was demonstrated recently by the so-called HERA anomaly
\cite{HERA}, in which the apparent excess of events at $x \sim 0.6$ and
$Q^2 \sim 30,000$ GeV$^2$, which triggered speculation about evidence
of leptoquarks, could be largely explained by a tiny modification in the
input valence distributions at $x \sim 0.8$ \cite{KLT,CHARM,CTEQ_LX}.

It is crucial therefore that a reliable method be found for extracting  
the free neutron structure function from measured cross sections.
While extracting $F_2^n$ from nuclear cross sections at large $x$ does
require knowledge of the nuclear EMC effect, it turns out that $F_2^n$
extracted from the ratio of deep inelastic $^3$He and $^3$H cross
sections is, within the likely experimental errors, almost completely
independent of the nuclear corrections.

% ........................................................................
\subsection{Extraction of $F_2^n$ from $A=3$ Mirror Nuclei}

Because the magnitude of the nuclear EMC effect increases with the
binding energy (or mass number $A$), light nuclei are naturally best
suited for playing the role of effective neutron targets.
Ideally, one should consider systems which maximize the symmetry between
the binding effects on the proton and neutron.
By comparing the effective `structure function' of a bound proton with
the free proton structure function $F_2^p$ (see Ref.~\cite{MST} for a
detailed discussion about the definition of bound nucleon structure
functions), one can infer the nuclear correction that must be applied
to obtain the free neutron $F_2^n$ from the bound neutron `structure  
function'.
Unfortunately, the lightest such system -- the deuteron -- is isoscalar,
so that the proton and neutron information cannot be separated through
inclusive scattering alone.

The three-nucleon system, on the other hand, offers a unique opportunity
for isolating the nuclear effects for both the bound proton and bound
neutron with totally inclusive scattering.
In a charge symmetric world the properties of a proton (neutron) bound
in a $^3$He nucleus would be identical to that of a neutron (proton)
bound in $^3$H.
If in addition the proton and neutron distributions in $^3$He (and in
$^3$H) were identical, the neutron structure function could be
extracted with {\em no nuclear corrections}, regardless of the size 
of the EMC effect in $^3$He or $^3$H separately.

In practice $^3$He and $^3$H are of course not perfect mirror nuclei
--- their binding energies for instance differ by $\sim 10\%$ ---
and the $p$ and $n$ distributions are not quite identical.
However, the $A=3$ system has been studied for many years, and modern
realistic $A=3$ wave functions are known to rather good accuracy.
In a self-consistent framework one can use the same $NN$ interaction to
describe the two-nucleon system ($NN$ scattering, deuteron form factors,
quasi-elastic $e D$ scattering, etc), as well as to provide the basic
input interaction into the three-nucleon calculation.
Therefore the wave functions can be tested against a large array of
observables which put rather strong constraints on the models.

We start by defining the EMC-type ratios for the $^3$He and $^3$H
structure functions (weighted by corresponding isospin factors):
\begin{subequations}
\begin{eqnarray}
R(^3{\rm He}) &=& { F_2^{^3{\rm He}} \over 2 F_2^p + F_2^n }\ , \\
R(^3{\rm H}) &=& { F_2^{^3{\rm H}} \over F_2^p + 2 F_2^n }\ .
\end{eqnarray}  
\end{subequations}%
The ratio of these,
\begin{eqnarray}
\label{rr}
{\cal R} &=& { R(^3{\rm He}) \over R(^3{\rm H}) }\ ,
\end{eqnarray}
can be inverted to yield the ratio of free neutron to proton
structure functions,
\begin{eqnarray}
\label{np}
{ F_2^n \over F_2^p }
&=& { 2 {\cal R} - F_2^{^3{\rm He}}/F_2^{^3{\rm H}}
\over 2 F_2^{^3{\rm He}}/F_2^{^3{\rm H}} - {\cal R} }\ .
\end{eqnarray}

If the neutron and proton distributions in the $A=3$ nuclei are not
dramatically different, one might expect ${\cal R} \approx 1$.
We stress that $F_2^n/F_2^p$ extracted from Eq.~(\ref{np}) does not
depend on the size of the EMC effect in $^3$He or $^3$H, but rather
only on the {\em ratio} of EMC effects in $^3$He and $^3$H.
In the following sections we show that while the variation in the
$A=3$ EMC effect can be up to 5\% at large $x$, the deviation from
unity of the ratio ${\cal R}$ is typically less than 1\%, and is
essentially independent of the model wave function.

%%%%%%%%%%%%%%%%%%%%%%%%%%%%%%%%%%%%%%%%%%%%%%%%%%%%%%%%%%%%%%%%%%%%%%%%%
\section{Deep Inelastic Scattering from $A=3$ Nuclei}

In this Section we outline the theoretical framework used to describe
deep inelastic structure functions from nuclei in terms of nucleonic
degrees of freedom.
Corrections to this approach will be discussed in Section~IV.

% .......................................................................
\subsection{Impulse Approximation}

The standard framework within which nucleon Fermi motion and binding
effects are described in deep inelastic scattering from a nucleus at
large $x$ ($x \agt 0.4$) is the nuclear impulse approximation, in which
the virtual photon scatters incoherently from individual nucleons in the
nucleon.
Earlier calculations of the EMC effect in $A=3$ nuclei within this
approach were reported in Ref.~\cite{US}.

The nuclear cross section is calculated by factorizing the
$\gamma^*$--nucleus interaction into $\gamma^*$--nucleon and
nucleon--nucleus amplitudes.
In the absence of relativistic and nucleon off-shell corrections
\cite{MST,GL,KMPW,MPT} (which for the deuteron were shown \cite{MSTD}
to be negligible, and which are also expected to be small for $A=3$),
the nuclear structure function can then be calculated by smearing the
nucleon structure function with a nucleon momentum distribution in the
nucleus \cite{CONV}.

Corrections to the impulse approximation appear in the guise of final
state interactions (interactions between the nucleon debris and recoil
nucleus remnants), multiple rescattering of the virtual photon from
more than one nucleon, as well as scattering from possible
non-nucleonic constituents in the nucleus.
The rescattering corrections are known to be important at small $x$,
giving rise to nuclear shadowing for $x \alt 0.1$ \cite{SHAD}, while
meson-exchange currents (at least for the case of the deuteron) give
rise to antishadowing at small $x$ \cite{KAPTAR,MTSHAD}.
Although there is strong evidence for a role for virtual $\Delta$'s in
{\it polarized} deep inelastic scattering on $^3$He \cite{BGST} there
is as yet no firm evidence of a role for non-nucleonic degrees of
freedom in unpolarized, nuclear deep inelastic scattering.

Within the impulse approximation (IA), in the region
$0.3 \alt x \alt 0.9$ the structure function $F_2^A$ of a nucleus with
mass number $A$ can be written (to order $\vec p \, ^2/M^2$ in the
nucleon momentum) as:
\begin{eqnarray}
\label{F2Afull}
F_2^A(x,Q^2) &=& \int d^4p \left( 1 + {p_z \over p_0} \right)
S(p)\ {\cal F}(p,Q^2)\ F_2^N(x/y,Q^2,p^2)\ ,
\end{eqnarray}
where $p$ is the momentum of the bound nucleon, $y=(p_0+p_z)/M$ is
the light-cone fraction of the nuclear momentum carried by the nucleon 
and $S(p)$ is the nucleon spectral function (see Section III.B below).
The kinematic factor ${\cal F}$ contains finite-$Q^2$ corrections
\cite{WEST},
\begin{eqnarray}
\label{fac}
{\cal F} &=& \left( 1 + \frac{4 M p_z x^2 r}{y Q^2} \right)^2
- \left( 2 \vec p \, ^2 - p_z^2 \right)
  \frac{r^2 x^2}{y^2 Q^2}\ ,
\end{eqnarray}
where $r = \nu/|{\vec q}| = 1/\sqrt{1 + 4 M^2 x^2/Q^2}$, $\nu$ and
$|{\vec q}|$ being the energy and the three-momentum transfer,
respectively, so that ${\cal F} \to 1$ as $Q^2 \to \infty$.
The function $F_2^N$ is the structure function of the bound (off-shell)
nucleon, which in general depends on the nucleon virtuality,
$p^2 \not= M^2$.
For non-relativistic systems, and away from the very large $y$ region,
the nucleon will not be very far off-shell, so that $F_2^N$ can be well
approximated by the free nucleon structure function (although in the
numerical results below we will consider the sensitivity of our results
to the $p^2$ dependence of $F_2^N$).
If $F_2^N$ is independent of $p^2$, one can factorize this from the
rest of the integrand in Eq.~(\ref{F2Afull}), which enables one to
write a simple convolution formula for the nuclear structure function,
\begin{eqnarray}
\label{convolution}
F_2^A(x,Q^2) &=& \int_x^A dy\ f(y,Q^2)\ F_2^N(x/y,Q^2)\
\equiv\ f \otimes F_2^N\ ,
\end{eqnarray}
where the function $f(y,Q^2)$ gives the light-cone distribution of
nucleons in the nucleus, and is related to the nucleon spectral
function, $S(p)$, by:
\begin{eqnarray}
f(y,Q^2) &=& \int d^4 p
\left( 1 + {p_z \over p_0} \right)
\delta \left( y - { p_0 + p_z \over M } \right) S(p)\ {\cal F}(p,Q^2)\ .
\end{eqnarray}
In the limit $Q^2 \to \infty$ the function $f(y,Q^2)$ reduces to
the familiar $Q^2$ independent function:
\begin{eqnarray}
f(y)
&=& 2 \pi M y \int_{E_{\rm min}} dE
\int_{p_{\rm min}(y,E)}^\infty d|\vec p\, |\ |\vec p\, |\ S(p)\ ,
\end{eqnarray}
where $E$ is the separation energy, and where the lower limit on the $p$
integration is given by \cite{CL}:
\begin{eqnarray}
p_{\rm min}(y,E) &=& \frac{1}{2}
\left| \frac{\zeta^2 + 2 M_{A-1}^* \zeta}{\zeta + M_{A-1}^*} \right|\ ,
\end{eqnarray}
with $\zeta = M (1-y) - E$ and $M_{A-1}^*$ is the mass of the (possibly
excited) residual nucleus.

The derivation of the impulse approximation expressions requires
knowledge of the struck nucleon's off-shellness, {\it i.e.} the
dependence of the nucleon structure function on the virtuality
of the struck nucleon.
Although a complete treatment of off-shell effects can only be given
within a fully relativistic description of nuclear dynamics, model
calculations exist which can estimate these corrections for DIS from
both the deuteron and complex nuclei.
Off-shell effects can be described within a formalism which introduces
corrections to the convolution formula of Eq.~(\ref{convolution}).
However, as explained below, although their influence is felt mostly at
large $x$, the ultimate effect on the extraction of the $F_2^n/F_2^p$
ratio from the ratio ${\cal R}$ is rather small.
(Note that some authors write the flux factor $(1 + p_z/p_0)$ in
Eq.~(\ref{F2Afull}) as $(1 + p_z/M)$ \cite{KPW}, or as $(p_0 + p_z)/M$
\cite{CL}.
To the order in which we work these are in fact equivalent,
and constitute small corrections numerically.)

A further simplification of Eq.~(\ref{convolution}) can be made by
observing that the nucleon momentum distributions $f(y)$ are strongly
peaked about $y=1$, so that by expanding the nucleon structure function
about this point one can obtain approximate expressions for the nuclear
structure functions in terms of average separation and kinematic
energies.
Keeping terms up to order $\vec p \, ^2/M^2$ (note that $E$ is of order
$\vec p \, ^2/2M$) one finds:
\begin{eqnarray}
\label{expansion}
F_2^A(x,Q^2) &\approx& 
F_2^N(x,Q^2)\
+\ x { \partial F_2^N(x,Q^2) \over \partial x }
        { \langle E \rangle + \langle T_R \rangle \over M}
+\ x^2 { \partial^2 F_2^N(x,Q^2) \over \partial x^2 }
        {\langle T \rangle \over 3 M}\ ,
\end{eqnarray}  
where
\begin{subequations}
\begin{eqnarray}
\langle E \rangle
&=& \int d^4p\ E\, S(p)\ ,                              \\
\langle T \rangle
&=& \int d^4p\ \frac{\vec p \, ^2}{2M}\ \, S(p)\ ,              \\
\langle T_R \rangle
&=& \int d^4p\ \frac{\vec p \, ^2}{2M_{A-1}^*}\ \, S(p)\ ,
\end{eqnarray}
\end{subequations}
are the average separation, kinetic and spectator recoil energies,
respectively.
Such an expansion will be useful in the next Section in identifying the
physical origin of the various contributions affecting the EMC ratios.
For example, as we discuss in Section II.E, the value of
$\langle E \rangle$ determines the position of the peak in the function
$f(y)$.

For the specific case of an $A=3$ nucleus the calculation of the nuclear
structure function amounts to determining the nucleon spectral function
from the three-body nuclear wave function.
The details are discussed in the next Section, where we present two
distinct and independent approaches, one by solving the homogeneous
Faddeev equation with a given two-body interaction \cite{BTA}, and the
other by using a variational technique \cite{CIOFI80,CIOFI84}.
In terms of the proton and neutron momentum distributions in $^3$He,
the nuclear structure function is given by:
\begin{eqnarray}
F_2^{^3{\rm He}}\
&=& 2\ f_{p/^3{\rm He}}\ \otimes\ F_2^p\
 +\    f_{n/^3{\rm He}}\ \otimes\ F_2^n\ .
\end{eqnarray}
Similarly for $^3$H, the structure function is evaluated from the proton
and neutron momentum distributions in $^3$H:
\begin{eqnarray}
F_2^{^3{\rm H}}
&=&    f_{p/^3{\rm H}} \otimes\ F_2^p\
 +\ 2\ f_{n/^3{\rm H}} \otimes\ F_2^n\ .
\end{eqnarray}
The proton and neutron distributions in $^3$H can be related to those
in $^3$He according to:
\begin{subequations}
\label{f3rels}
\begin{eqnarray}
f_{n/^3{\rm H}} &=& f_{p/^3{\rm He}}\ +\ \Delta f_p\
                \equiv\ f_p\ +\ \Delta f_p\ ,   \\
f_{p/^3{\rm H}} &=& f_{n/^3{\rm He}}\ +\ \Delta f_n\
                \equiv\ f_n\ +\ \Delta f_n\ .
\end{eqnarray}
\end{subequations}%
Because charge symmetry breaking effects in nuclei are quite small,
one can usually assume that $\Delta f_p \approx \Delta f_n \approx 0$,
although in practice we consider both charge symmetric and charge
symmetry breaking cases explicitly.

% .......................................................................
\subsection{Three-Nucleon Spectral Function}

Calculations of the structure functions of $A=3$ nuclei can be performed
by using realistic three-body spectral functions.  
In this Section we first describe the relevant features of the spectral
functions which determine the behavior of nuclear effects in DIS,
following which we outline two different methods of computing the
three-nucleon wave function, namely, via the Faddeev equations
\cite{BTA,FADDEEV,GIBSON} and the variational approach
\cite{CIOFI80,CIOFI84}.

To simplify the problem both theoretically and numerically, we will in  
the first instance consider the three-nucleon system with exact charge
symmetry, so that both the $^3$H and $^3$He wave functions can be
calculated simultaneously.
The Coulomb interaction will of course modify the wave functions
slightly through explicit charge symmetry breaking effects, giving rise
to the difference between $^3$H and $^3$He binding energies.
We subsequently examine the effects of the binding energy on the
structure functions.

The models we consider are based on two-body interactions.
Possible three-body forces do not provide any significant improvement
in the quality of the results, and are considerably more
difficult to take into account.
For the charge-symmetric case one can treat $^3$He and $^3$H as members
of an exact isospin doublet.

The nucleon spectral function is the joint probability of finding
a nucleon in the nucleus $A$, with three-momentum $\vec p$ and removal
energy $E$.
If at the values of momentum and energy transfer considered the outgoing
nucleon's motion is described by a plane wave, the spectral function can
be written as the sum of the momentum densities for each final state:  
\begin{eqnarray}
S(p) &=&
\frac{1}{(2 \pi)^3} \sum_f
\left| \int d^3 r \, e^{i {\vec p} \cdot {\vec r}}
        G_{f o}({\vec r})
\right|^2
\delta(E - (E_2^f - E_3) )\ ,
\label{spectral}
\end{eqnarray}
where $E_2^f$ and $E_3$ are the values of the total energy of the the
two-nucleon spectator system and of the initial nucleus, respectively;
$G_{f o}({\vec r})$ is the overlap between the initial and final wave
functions in coordinate space, with the $A-1$ (spectator) system being
described in terms of a complete set of final states.
The spectral function is normalized according to:
\begin{eqnarray}
\int d^4p\ S(p) &=& 1\ .
\end{eqnarray}
Integrating the spectral function over the energy defines the nucleon
momentum distribution in the nucleus:
\begin{eqnarray}
\int dE\ S(p,E) &=& n(p)\ .
\end{eqnarray}
There are in general two processes which can contribute to deep
inelastic scattering from $^3{\rm He}$:
{\it i)} two-body breakup (with a deuteron, $d$, in the final state) and
{\it ii)} three-body breakup, $(pn)$ and $(pp)$;
analogously, for $^3{\rm H}$ one has:
{\it i)} two-body breakup ($d$) and
{\it ii)} three-body breakup, $(np)$ and $(pp)$.

We write the spectral functions for the two nuclei, distinguishing
between scattering from proton and neutron, as:
\begin{subequations}
\begin{eqnarray}
S_{^3{\rm He}}(p)
&=& \frac{2}{3} S_{p/^3{\rm He}}(p)\
 +\ \frac{1}{3} S_{n/^3{\rm He}}(p)\ ,          \\  
S_{^3{\rm H}}(p)
&=& \frac{1}{3} S_{p/^3{\rm H}}(p)\
 +\ \frac{2}{3} S_{n/^3{\rm H}}(p)\ ,
\end{eqnarray} 
\end{subequations}%
where, in analogy with Eq.~(\ref{f3rels}), the proton and neutron
spectral functions in $^3{\rm He}$ and $^3{\rm H}$ are related by:
\begin{subequations}
\begin{eqnarray}
S_{p/^3{\rm He}}(p)
&=& S_{n/^3{\rm H}}(p) \equiv S_p(p)\ +\ \Delta S_p(p)\ ,       \\
S_{n/^3{\rm He}}(p)
&=& S_{p/^3{\rm H}}(p) \equiv S_n(p)\ +\ \Delta S_p(p)\ ,
\end{eqnarray}
\end{subequations}
with the terms $\Delta S_{p,n}(p)$ representing explicit isospin symmetry
breaking corrections.

By breaking down the spectral functions into contributions
corresponding to two-body and three-body final states, on has
\begin{subequations}
\begin{eqnarray}
S_p(p) &=& S_p^{(2)}(p) + S_p^{(3)}(p)\ ,               \\
S_n(p) &\equiv & S_n^{(3)}(p)\ ,
\end{eqnarray}
\end{subequations}%
where $S_p^{(2)}$ and $S_p^{(3)}$ represent the contributions to the proton
spectral function from a deuteron and $(np)$ break up final states, while
for the neutron spectral function only the $pp$ final state contributes.
In terms of these components, the average separation and kinetic energies
can be written as:
\begin{subequations}
\begin{eqnarray}
\langle E \rangle
&=& \frac{2}{3} \int d^4p\
    \left( S_p^{(2)}(p)\ +\ S_p^{(3)}(p) \right)\ E\
 +\ \frac{1}{3} \int d^4p\ S_n(p)\ E                    \nonumber\\
&=& \frac{2}{3}  \langle E_p^{(2)} \rangle  
 + \frac{1}{3} \left( 2 \langle E_p^{(3)} \rangle + \langle E_n \rangle
               \right)\ ,
                        \\
\langle T \rangle
&=& \frac{2}{3} \int d^4p\
    \left( S_p^{(2)}(p)\ +\ S_p^{(3)}(p) \right)\ \frac{\vec p \, ^2}{2M}
 +\ \frac{1}{3} \int d^4p\ S_n(p) \frac{\vec p \, ^2}{2M}  \nonumber\\
&=& \frac{2}{3} \langle T_p^{(2)} \rangle  +
        \frac{1}{3} \left( 2 \langle T_p^{(3)} \rangle
        +\ \langle T_n \rangle \right)\ .
\end{eqnarray}
\end{subequations}%
The normalization of the spectral function is written in terms of
the normalizations for the two-body and three-body breakup spectral
functions, ${\cal N}_p^{(2)}$ and ${\cal N}_p^{(3)}$, as:
\begin{eqnarray}
\label{Scmpts}
1 &=& \frac{2}{3} \int d^4p\ \left( S_p^{(2)}(p) + S_p^{(3)}(p) \right)\
   +\ \frac{1}{3} \int d^4p\ S_n(p)\
   =\ \frac{2}{3} \left( {\cal N}_p^{(d)} + {\cal N}_p^{(np)} \right)
   +\ \frac{1}{3}\ .
\end{eqnarray}

In summary: we have shown the features of the spectral function, $S(p)$
and of the light-cone function $f(y)$, which determine the behavior of
the nuclear corrections to the deep inelastic structure functions at
$x \ge 0.2$. 
While details of the short range structure could be important in 
determining the behavior at very large $x$, for $x \leq 0.6$--0.7 the
nuclear modifications are determined by the values of the average
removal and kinetic energies and therefore only loosely related to
the detailed structure of the spectral function.
Thus, we can safely state that nuclear effects are under control.

Having developed the formalism, in the following we describe the
evaluation of the spectral function, within the Faddeev and
variational approaches, from which the nuclear structure function
will be calculated.

% . . . . . . . . . . . . . . . . . . . . . . . . . . . . . . . . . . . .
\subsubsection{Faddeev Equations}

A full description of the calculation of the Faddeev wave function used
here has been given in Ref.~\cite{BTA}.
We therefore only briefly outline the calculation here.
We work in momentum space using a separable potential, which further
simplifies the computation \cite{SZ}.
The wave function is written as a sum of so-called ``Faddeev
components''\cite{AB,LEV}:
\begin{equation}
\label{fadcomp1}
  \left\vert\Psi\right\rangle
= \left\vert\varphi _\alpha \right\rangle
+ \left\vert\varphi_\beta \right\rangle
+ \left\vert\varphi_\gamma \right\rangle
= \{e+(\alpha\beta\gamma)+(\alpha\gamma\beta)\}\left\vert\varphi_\gamma
  \right\rangle\ ,
\end{equation}
where $\alpha$, $\beta$ and $\gamma$ are indices running from 1 to 3
(with $\alpha \not= \beta \not= \gamma$).
In this equation ``$e$'' is the neutral element of the permutation
group of three objects, and ``$(\alpha\beta\gamma)$'' and
``$(\alpha\gamma\beta)$'' are cyclic permutations.
$\left\vert\varphi _\alpha \right\rangle$ is referred to as the
``Faddeev component'' of the wave function in which the spectators to
the nucleon $\alpha$ interact last \cite{FAD_NOTE}.

Using the symmetry properties of the wave function (see e.g.
Refs.~\cite{BTA,LEV}), one writes a set of coupled equations for the
Faddeev components:
\begin{equation}
\label{phia}
\left\vert\varphi_\alpha \right\rangle
= G_0 t_\alpha 
( \left\vert\varphi_\beta \right\rangle
+ \left\vert\varphi_\gamma \right\rangle )\ ,
\end{equation}
where $t_\alpha$ is the usual $t$-matrix defined by the 
Lippmann-Schwinger equation:
\begin{equation}
\label{deftmatrix}
t_\alpha (E) = 
V_\alpha +V_\alpha G_0 (E)t_\alpha (E)=(1-G_0 (E)V_\alpha)^{-1}V_\alpha\ ,
\end{equation}
with $G_0 (E)=(E-H_0)^{-1}$ and $V_\alpha$ the interaction between
particles $\beta$ and $\gamma$.
{}From these expressions one can derive a set of homogeneous Faddeev
equations for the spectator function $\Xi$ \cite{FBS11-89},
\begin{equation}
\label{eqXimomentum}
\Xi_{_{N_\alpha}}(p_\alpha )
= 2\sum _{N^{\prime}_\alpha N^{}_\beta}
  \tau_{_{N^{}_\alpha N^{\prime}_\alpha}}(E,p_\alpha )\int ^{+\infty}_0
  dp^{}_\beta\ p^{2}_\beta\
  {\cal Z}_{_{N^{\prime}_\alpha N_\beta}}(E,p_\alpha ,p_\beta)\
  \Xi_{_{N_\beta}}(p_\beta )\ ,
\end{equation}
where ${\cal Z}$ is the kernel of the integral equation, and the
matrix $\tau_{_{N^{}_\alpha N^{\prime}_\alpha}}$ is related
to the $t$-matrix by:
\begin{equation}
\label{septmatrix}
t_{n^{}_\alpha n^{\prime}_\alpha}(E)
= \left\vert g_{n^{}_\alpha}\right\rangle
  \tau_{n^{}_\alpha n^{\prime}_\alpha}(E)\left\langle
  g_{n^{\prime}_\alpha}\right\vert\ .
\end{equation}
Here a three-nucleon channel is denoted by an index $N_\alpha$ and a
two-nucleon channel by an index $n_\alpha$.
The form factor $g_{n_\alpha}$ is defined by the form of the separable
potential.
Details of the computation of the $\tau$-matrix and the kernel are
given in Refs.~\cite{LEV,AT} and \cite{BTA,AT,AT3}, respectively.

The relevance of the spectator function becomes clear if one considers
the relation between $\Xi$ and $\varphi_\alpha$:
\begin{equation}
\label{eqeta1c}
\left\langle\Omega_{N_\alpha}^{JI}\big\vert\varphi_\alpha\right\rangle
= 2G_0 (E)\left\vert g_{_{N^{}_\alpha}}\right\rangle
\left\vert\Xi_{_{N^{}_\alpha}}\right\rangle\ ,
\end{equation}
where $\left\vert\Omega_{N_\alpha}^{JI}\right\rangle$ is the angular
element of our partial wave decomposition for isospin $I$ and spin $J$.
The homogeneous equation (\ref{eqXimomentum}) then enables one to
compute the contribution from one of the Faddeev components to the
total wave function.
The total wave function relative to the decomposition in the 
$\left\vert\Omega_{N_\alpha}^{JI}\right\rangle$ 
partial wave also requires the contributions
$\left\langle\Omega_{N_\alpha}^{JI}\big\vert\varphi_\beta\right\rangle$
and
$\left\langle\Omega_{N_\alpha}^{JI}\big\vert\varphi_\gamma\right\rangle$.
Since one has a system of identical particles, these two contributions
are equal for obvious reasons of symmetry.
Details of the computation of this contribution can be found in 
Ref.~\cite{BTA}.

To examine the model dependence of the distribution function we use
several different potentials, namely the ``EST'' (Ernst-Shakin-Thaler)
separable approximation to the Paris potential \cite{PEST} (referred to
as ``PEST''), the unitary pole approximation \cite{SA} to the Reid Soft
Core (RSC) potential \cite{RSC}, and the Yamaguchi potential \cite{YAM}
with 7\% mixing between $^3 S_1$ and $^3 D_1$ waves.
The homogeneous Faddeev equation was solved with 5 channels for both
potentials.
The results for the tri-nucleon binding energies are --7.266MeV (PEST)
and --8.047MeV (Yamaguchi), which differ by $\sim 14\%$ and $\sim 5\%$,
respectively, from the experimental $^3$H binding energy of --8.482MeV
(one expects the binding energy from this tri-nucleon calculation to be
closer to the experimental $^3$H binding energy than $^3$He since one
does not expect Coulomb corrections for $^3$H).

The issue of the binding energy is well known, and this result is
consistent with what one usually expects when the Coulomb interaction
is switched off.
To estimate the effect of neglecting the Coulomb interaction in $^3$He
and at the same time correct the long range part of the three-body wave
function due to the change in the binding energy, we have modified the 
$^1S_0$ potential in $^3$He and $^3$H to reproduce their respective  
experimental energies.
This leaves the $^3S_1$--$^3D_1$ interaction responsible for the
formation of the deuteron unchanged, and introduces rather strong
charge symmetry breaking in the system.
This approximation distributes the symmetry breaking effects of the
Coulomb interaction equally over the three particles, whereas in the
exact case it should only arise from the difference between $p p$ and
$n p$ interactions.
It therefore represents an overestimate of any charge symmetry breaking
effects since one attributes to charge symmetry breaking an effect which
should partly come from three-body forces.
However, this simple modification to the $^1S_0$ interaction will allow
us to study explicitly the possible effects on the deep inelastic
structure functions associated with the differences in the binding
energies of $^3$He and $^3$H.

% . . . . . . . . . . . . . . . . . . . . . . . . . . . . . . . . . . . .
\subsubsection{Variational Approach}

In the variational approach one writes the overlap integral in coordinate
space, $G_{f o}({\vec r})$, Eq.~(\ref{spectral}), as:
\begin{equation}
G_{f o}({\vec r})
 = {\cal N} \int d^3 {\vec \rho} \;
   \psi^f_2({\vec \rho})\ \psi^i_3({\vec r},{\vec \rho})\ ,
\end{equation}
where ${\cal N}$ is a normalization factor; $\psi^f_2({\vec \rho})$ and
$\psi^i_3({\vec r},{\vec \rho})$ are the wave functions with eigenvalue
$E_2^f$ for the spectator two-body system, and with eigenvalue $E_3$ for
the initial three-body system, respectively;
$f \equiv (J_f, M_f, S_f, \lambda)$ represents the quantum numbers of the
spectator system, $\lambda$ specifying the tensor coupled states at high
energy and momentum, and $i \equiv (1/2, M)$;
${\vec r}$ and ${\vec \rho}$ are the intrinsic coordinates for the three
body system \cite{VAR_NOTE}.

The three-body wave function is found by diagonalization of the intrinsic
nuclear Hamiltonian using an L-S coupling scheme, and the basis
\begin{equation}
\left| \phi_K \right. \rangle =  
\left. \left| (L {\it l}) {\it L}, \left( S \frac{1}{2} \right) {\it S};
        \frac{1}{2} M \right. \right\rangle\ ,
\end{equation}
where $L$ and $l$ refer to two sets of harmonic oscillator wave
functions with different harmonic oscillator parameters \cite{NUN}. 
The wave function is then written schematically as:
\begin{equation}
\psi^i_3({\vec r,\rho}) = \sum_K \left| \phi_K \right. \rangle\ ,
\end{equation}
where the relevant components are the ones with ${\it L}=0$ and
${\it L}=2$.
All calculations using the variational method outlined here have been
performed using the RSC \cite{RSC} interaction.

The two-body spectator wave function describes either a deuteron, 
$\psi^f_2({\vec r}) \equiv \psi_d({\vec r})$ (two-body channel), 
or an interacting nucleon pair (three-body channel).
The corresponding quantum numbers and ground state energy values are:
$f \equiv (1,M_J,1)$ and $E_2^f=-2.23$ MeV (two-body channel);
$f \equiv (J,M_J,S,\lambda)$ and $E_2^f > 0$ (three-body channel). 
The three-body channel wave function calculated in \cite{CIOFI84}, 
considers states up to $J=5$, using the RSC interaction up to $J = 2$.
For higher values of $J$ the interaction among the two nucleons is
assumed to be negligible.

Analogous issues as for the Faddeev calculations outlined above are
present in the variational approach, namely, discrepancies in the
theoretical values of the binding energy depending on the type of
potential, the accurate handling of Coulomb effects, and the possible
presence of charge symmetry breaking effects.         
These issues are examined quantitatively in Section~IV.

% .......................................................................
\subsection{Nucleon Momentum Distributions}

Before proceeding to the evaluation of the structure functions in terms
of the nuclear spectral functions, we first review some general
features of the spectral functions and 
light-cone momentum distribution, $f(y)$.

The relevant features of the $^3$He spectral function are:

\noindent {\it i)}
A pronounced peak at $E=2$ MeV, corresponding to the case in which the
spectator deuteron recoils.

\noindent {\it ii)}
Some strength extending to high values of the energy and momentum
($p \gtrsim 300$ MeV) but lying at least three orders of magnitude
below the peak.
The high momentum and energy part of the spectral function is given
almost entirely by the short range part of the nucleon--nucleon
interaction which is actually responsible for break-up configurations
of the spectator system. 
In heavier nuclei these components can be calculated using two-nucleon
correlations, as described by Ciofi~degli~Atti {\em et al.} \cite{CORREL}.

The function $f(y)$ reflects the features of the spectral function
described above.
Namely, it has a sharp peak in the vicinity of $y=1$,
$y_{\rm peak} \approx 1 - \langle E \rangle/M$ (modulo spectator recoil
corrections, see below), and some strength away from $y_{\rm peak}$ is
present which integrates to a considerable fraction of the total
strength.
For the proton, all the distributions have a similar shape and peak
value, however, for the neutron the variational distribution peaks at
slightly smaller $y$ and has a larger tail than the Faddeev.
The origin of this is the larger momentum components in the deuteron
spectator part of the neutron distribution in the variational
distribution than in the corresponding Faddeev distribution.

The main contribution of $f(y)$ in the convolution formula is from its
values around $y_{\rm peak}$, namely one can write:
\begin{equation}
F_2^A (x,Q^2) \approx F_2^N(x/y_{peak},Q^2) < F_2^N(x,Q^2)\ .
\label{depletion}
\end{equation}
Since $F_2^N$ is a decreasing function of $x$ in the interval
$0.2 \lesssim x \lesssim 0.6 $, this gives rise to the depletion
in the EMC ratio, $F_2^A/F_2^N$.
At larger $x$, the EMC ratio rises above unity because of the different
kinematic boundaries affecting the smearing, namely using the
asymptotic convolution formula, the kinematic thresholds for the free
nucleon, the deuteron and $A=3$ nuclei are located at $x=1$, $x=2$ and
$x=3$, respectively.

In summary, the EMC effect at intermediate values of $x$
($0.2 \leq x \leq 0.65$) is determined almost entirely by the average
values of the removal and kinetic energies, Eq.~(\ref{depletion}) and
Eq.~(\ref{expansion}).
At larger values of $x$, the approximations Eq.~(\ref{depletion}) and
Eq.~(\ref{expansion}) start breaking down, and the EMC effect is
directly sensitive to the large energy and momentum components of the
spectral function.

Note also that $f(y)$ can be translated easily into the ``Y-scaling''
function $F(Y)$ \cite{YSCL}, extracted from quasi-elastic scattering.
The variable $Y$ is given in terms of $y$ and the nucleon and nuclear
masses as:
$Y= (1/2) [M_A-My)^2-M_{A-1}^2]/(M_A-My)$, which allows one to relate
$F(Y)= f(y)/M$.
Unlike in DIS, the nuclear cross sections for quasi-elastic scattering
are given directly in terms of $f(y)$, so that quasi-elastic data can
be used in addition to constrain models of nuclear dynamics.
A quantitative description, however, of quasi-elastic scattering
requires additional contributions beyond the impulse approximation,
such as from meson-exchange currents, which do not contribute in deep
inelastic scattering.
In our analysis we use distributions which are consistent with those
used in standard analyses of the quasi-elastic scattering data.

% .......................................................................
\subsection{EMC Effect in $A=3$ Nuclei}

Before proceeding to the calculation of the ratio ${\cal R}$ of the
EMC effects in $^3{\rm He}$ and $^3{\rm H}$, and the associated
sensitivity of the extracted $F_2^n/F_2^p$ to ${\cal R}$, we first
discuss the predictions of the conventional nuclear models for the
absolute EMC ratios and compare with available data.

As well as offering a relatively clean way to extract $F_2^n$ from
nuclear data, the $A=3$ system is also a valuable laboratory for
testing models of the EMC effect for few-body nuclei.
Although the determination of $F_2^n/F_2^p$ requires only the ratio of
$^3{\rm He}$ to $^3{\rm H}$ structure functions, data on the absolute
values of $F_2^{^3{\rm He}}$ and $F_2^{^3{\rm H}}$ can in addition fix
the magnitude of the EMC effect in $A=3$ nuclei:
\begin{subequations}
\begin{eqnarray}
R(^3{\rm He}) &=&
{ F_2^{^3{\rm He}} \over
  F_2^p \left( 2 + \left. F_2^n/F_2^p\right|_{\rm extr} \right) }\ , \\
R(^3{\rm H}) &=&
{ F_2^{^3{\rm H}} \over
  F_2^p \left( 1 + 2 \left. F_2^n/F_2^p\right|_{\rm extr} \right) }\ .
\end{eqnarray}
\end{subequations}%
Unfortunately, at present there are no data at all on the
$F_2^{^3{\rm H}}$ structure function, and only scant information on
$F_2^{^3{\rm He}}$, from a recent HERMES measurement \cite{HERMES},
the main focus of which was the low $x$, low $Q^2$ region.
Nevertheless, the available data can provide a useful check on the
calculation.

In Fig.~2 the ratio of the $^3$He to free nucleon (corrected for
non-isoscalarity) structure functions is shown for the Faddeev (PEST)
and variational wave functions, compared with the HERMES data
\cite{HERMES} on the ratio of $\sigma(^3{\rm He})/(\sigma(d)+\sigma(p))$.
The difference between the solid and dashed curves in Fig.~2
illustrates the effect on the ratio due to possible nuclear corrections
in deuterium.
The various models predict qualitatively similar behavior for the
ratio as a function of $x$, with the magnitude of the depletion at
$x \sim 0.5$--0.7 ranging from $\sim$ 2\% in the variational approach
to $\sim$ 4\% using the Faddeev wave functions.
Within the relatively large errors for $x \agt 0.4$, the agreement
between the models and the experiment is reasonably good.

Similar behavior is found for the ratio of $^3$H to isoscalar nucleon
structure functions, illustrated in Fig.~3 for the Faddeev and
variational calculations.
The trough at $x \sim 0.6$ in $^3$H is predicted to be slightly deeper
than that in $^3$He in all models.
The dependence on the input potential is negligible, as the PEST and
RSC Faddeev results illustrate.
Data on $^3$H, and better quality data extending to larger $x$ for
$^3$He, would clearly be of great value in constraining models of the
EMC effect in $A=3$ nuclei.

%%%%%%%%%%%%%%%%%%%%%%%%%%%%%%%%%%%%%%%%%%%%%%%%%%%%%%%%%%%%%%%%%%%%%%%%%
\section{Ratio of Ratios}

In this Section we discuss the model dependence of the ratio ${\cal R}$
of the $^3$He and $^3$H EMC ratios arising from uncertainty in the
nuclear wave function, the off-shell modifications of the nucleon
structure function, and possible non-nucleonic degrees of freedom in
the $A=3$ nuclei.
While the magnitude of the EMC effect in $^3$He and $^3$H was found in
the previous Section to differ by as much as several percent at
$x \lesssim 0.8$ in different models, one expects the ratio of these
to be considerably less model dependent.

% .......................................................................
\subsection{Nuclear Wave Function Dependence}

Using the light-cone momentum distributions described in Section~III,
the ratio ${\cal R} = R(^3{\rm He})/R(^3{\rm H})$ of EMC ratios for
$^3$He to $^3$H is shown in Fig.~4 for various nuclear model wave
functions, namely, Faddeev with the PEST, RSC and Yamaguchi potentials,
and variational using the RSC potential.
(Unless otherwise stated, in all cases the CTEQ5 parameterization
\cite{CTEQ} of parton distributions at $Q^2=10$~GeV$^2$ will be used
for $F_2^N$.)
The EMC effects are seen to mostly cancel over a large range of $x$,
out to $x \sim 0.8$, with the deviation from a `central value'
${\cal R} \approx 1.01$ within $\pm 1\%$.
The larger absolute EMC effects in $^3$He and $^3$H predicted with the
Faddeev calculations in Figs.~2 and 3 are reflected in a larger
deviation of ${\cal R}$ from unity than with the variational wave
functions, as seen in the three Faddeev calculations in Fig.~4.
Furthermore, the dependence on the $NN$ potential is very weak.
In practice, the exact shape of ${\cal R}$ will not be important for
the purposes of extracting $F_2^n/F_2^p$ from the
$F_2^{^3{\rm He}}/F_2^{^3{\rm H}}$ ratio; rather, it is essential that
the model dependence of the deviation of ${\cal R}$ from the central
value should be small.

% .......................................................................
\subsection{Charge Symmetry Breaking}

The ratio ${\cal R}$ in Fig.~4 was calculated using three-nucleon wave
functions neglecting the Coulomb interaction and working in an isospin
basis \cite{3BODY}.
To estimate the effect of neglecting the Coulomb interaction in $^3$He
and at the same time correct the long range part of the three-body wave
function due to the change in the binding energy, we modify the $^1S_0$
potential in the $^3$He and $^3$H to reproduce their respective
experimental energies.
In this way the $^3S_1-^3D_1$ interaction responsible for the formation
of the deuteron is unchanged.
This approximation spreads the effect of the Coulomb interaction over
both the $pp$ and $np$ interaction in the $^1S_0$ channel.
To that extent, it shifts some of the Coulomb effects in the neutron
distribution in $^3$He to the proton distribution.
However, this simple modification to the $^1S_0$ interaction allows one
to study explicitly the possible effects associated with the differences
in the binding energies of $^3$He and $^3$H.

The ratio ${\cal R}$ calculated with the Faddeev (PEST) wave function
modified according to this prescription is shown in Fig.~5 (dashed
curve), compared with the charge symmetric result (solid).
The effect of this modification is a shift of approximately
$\lesssim 0.5\%$ in ${\cal R}$, maximal at $x \sim 0.65$.
The effects of charge symmetry breaking therefore still leave a ratio
which deviates from unity by $\lesssim 2\%$.

% .......................................................................
\subsection{Finite $Q^2$ Effects}

The structure function ratios discussed above are calculated assuming
leading twist dominance of the nucleon structure function at
$Q^2=10$~GeV$^2$.
At finite $Q^2$ there will be contributions from both the kinematic
factor ${\cal F}$ in the inelastic structure function,
Eq.~(\ref{F2Afull}), and from higher twist ($\propto 1/Q^2$) effects,
including quasi-elastic scattering from the bound nucleon, which may
not be negligible at large $x$ \cite{CDL}.
To illustrate the impact of these finite $Q^2$ effects on the ratio
${\cal R}$, in Fig.~6 we show the ratio at several values of $Q^2$
($Q^2=4$~GeV$^2$ and 20~GeV$^2$), together with the asymptotic result
(${\cal F} \to 1$).
To facilitate the comparison, all curves have been obtained using the
variational $A=3$ wave functions.
The points denoted by bullets correspond to values of $x$ and $Q^2$
that would be relevant for kinematics at a 12~GeV Jefferson Lab
facility \cite{WKSHP} (see Section~V and Table~I below), for which
$Q^2$ varies from $Q^2=3$~GeV$^2$ in the lowest $x$ bin, to
$Q^2=14$~GeV$^2$ at the highest ($x \approx 0.8$).
The effect of the $Q^2$ dependence is clearly rather modest.

The role of quasi-elastic scattering is illustrated by the dashed
curve in Fig.~6, for $Q^2=4$~GeV$^2$.
For $x \lesssim 0.8$ the quasi-elastic contribution is negligible
for the relevant kinematics, with a correction of order 1\% at
$x = 0.8$.
At the largest values of $x$, for instance, where $Q^2=14$~GeV$^2$,
we have checked that the quasi-elastic contribution is suppressed.
Its effect does start to become important, however, for $x \gtrsim 0.85$
at fixed $Q^2 \lesssim 5$~GeV$^2$, as can be seen from the wiggle
produced in the dashed curve in Fig.~8.

To test the sensitivity of ${\cal R}$ to higher twist corrections, we
compute the ratio using the fit to the total $F_2$ structure function
from Ref.~\cite{DOLA}, which includes both leading and subleading
effects in $1/Q^2$.
The difference between the leading twist only and leading $+$ higher
twist curves, represented by the lower and upper dashed curves in Fig.~7
(``DL'' and ``DL(HT)'' respectively), is negligible for $x \alt 0.8$,
increasing to $\sim 1\%$ at $x \sim 0.85$.
The size of the higher twist corrections can be determined by taking
measurements at several values of $Q^2$ and observing any $1/Q^2$
dependence of the structure function.
In particular, since the $Q^2$ dependence of $F_2^p$ has been measured
in a number of earlier experiments \cite{HT}, the $Q^2$ dependence of
the extracted $F_2^n/F_2^p$ ratio can be used to separate the leading
twist from the non-leading twist components of $F_2^n$ \cite{AKL}.

% .......................................................................
\subsection{Iteration Procedure}

The dependence of ${\cal R}$ on different input nucleon structure
function parameterizations is illustrated in Fig.~7, where several
representative parton distribution function fits are given at
$Q^2 = 10$~GeV$^2$.
Apart from the standard CTEQ fit (solid), the results for the GRV
\cite{GRV} (dot-dashed), Donnachie-Landshoff (DL) \cite{DOLA} (dashed),
and BBS \cite{BBS} (dotted) parameterizations are also shown
(the latter at $Q^2=4$~GeV$^2$).
For $x \alt 0.6$ there is little dependence ($\alt 0.5\%$) in the ratio
on the structure function input.
For $0.6 \alt x \alt 0.85$ the dependence is greater, but still with
$\alt \pm 1\%$ deviation away from the central value
${\cal R} \approx 1.01$.
The spread in this region is due mainly to the poor knowledge of the
neutron structure function at large $x$.
Beyond $x \approx 0.85$ there are few data in the deep inelastic region
on either the neutron or proton structure functions, so here both the
$d$ and $u$ quark distributions are poorly determined.

A standard assumption in most global fits of parton distributions is
that $d/u \to 0$ as $x \to 1$.
This assumption has recently been questioned on theoretical and
phenomenological grounds \cite{MT,W}.
The BBS parameterization \cite{BBS}, on the other hand, incorporates
constraints from perturbative QCD, and forces $d/u \to 0.2$ as $x \to 1$
\cite{FJ}.
The effect of the different large-$x$ behavior of the $d$ quark is  
apparent only for $x \agt 0.85$, where it gives a difference of
$\sim$ 1--2\% in ${\cal R}$ compared with the fits in which $d/u \to 0$.
One can also modify the standard CTEQ fit, for example, by applying a
correction factor \cite{W} to enforce $d/u \to 0.2$. However, this also
produces differences in ${\cal R}$ which are $\alt 2\%$ for $x < 0.9$.

Despite the seemingly strong dependence on the nucleon structure
function input at very large $x$, this dependence is actually
artificial.
In practice, once the ratio $F_2^{^3{\rm He}}/F_2^{^3{\rm H}}$ is
measured, one can employ an iterative procedure to eliminate the 
dependence altogether \cite{AFNAN,PSS,DECONV}.
Namely, after extracting $F_2^n/F_2^p$ from the data using some
calculated ${\cal R}$, the extracted $F_2^n$ can then be used to 
compute a new ${\cal R}$, which is then used to extract a new and
better value of $F_2^n/F_2^p$.
This procedure is iterated until convergence is achieved and a  
self-consistent solution for the extracted $F_2^n/F_2^p$ and
${\cal R}$ is obtained.
The results of this procedure are shown in Fig.~8 for different numbers
of iterations using as input $F_2^n/F_2^p = 1$.
The convergence is relatively rapid --- by the third iteration the
extracted function is almost indistinguishable from the exact result.
Although the effect on ${\cal R}$ from the present lack of knowledge
of the nucleon structure function is $\alt 2\%$ for $x \alt 0.85$, this
uncertainty can in principle be eliminated altogether via iteration, so
that the only model dependence of ${\cal R}$ will be from the nuclear
interaction in the $A=3$ nucleus.

Of course the accuracy of the iteration procedure is only as good as
the reliability of the formalism in Sec.~III used to calculate the
nuclear structure functions allows.
As pointed out in Ref.~\cite{SSS}, large corrections to the smearing
expression (\ref{F2Afull}) could lead to inaccuracies in the extracted
$F_2^n/F_2^p$ ratio.
In particular, it was argued \cite{SSS} that strong isospin-dependent
off-shell effects could give significantly larger deviations of
${\cal R}$ from unity than that found in Refs.~\cite{AFNAN,PSS}.
In the following we shall carefully examine the issue of off-shell
effects in $A=3$ nuclei and their effect on the ${\cal R}$ ratio.

% .......................................................................
\subsection{Nucleon Off-Shell Deformation}

The derivation of the convolution approximation in
Eq.~(\ref{convolution}) assumes that the nucleon off-shell dependence
in the bound nucleon structure function in Eq.~(\ref{F2Afull}) is
negligible.
In this Section we examine the accuracy of this assumption.
The off-shell dependence of $F_2^N$ is, as a matter of principle, not
measurable, since one can always redefine the nuclear spectral function
to absorb any $p^2$ dependence in the bound nucleon structure function.
However, off-shell effects can be identified once a particular form of
the interaction of a nucleon with the surrounding nuclear medium is
specified.
The discussion of off-shell modification of the nucleon structure
function in the nuclear medium is therefore understood to be within the
framework of the nuclear spectral functions defined in Section~III.

In convolution models off-shell corrections can arise both
kinematically, through the transverse motion of the nucleon in the
nucleus, and dynamically, from modifications of the bound nucleon's
internal structure.
Kinematical off-shell effects are essentially model independent, as
discussed in Ref.~\cite{GL}, while dynamical off-shell effects do depend
on descriptions of the intrinsic deformation of the bound nucleon
structure and are therefore model dependent.
The latter have been modeled for instance in a covariant spectator model
\cite{MST}, in which the DIS from a bound nucleon is described in terms
of relativistic vertex functions which parameterize the
nucleon--quark--spectator ``diquark'' interaction.
The dependence of the vertex functions on the quark momentum and the
``diquark'' energy is constrained by fitting to the on-shell nucleon
(proton) structure function data, while the additional dependence on
the virtuality of the off-shell nucleon can be constrained by comparing
the calculated nuclear structure function with the inclusive $F_2^A$
data.

Taking the nucleon's off-shellness into account, the bound nucleon
structure function in Eq.~(\ref{convolution}) can be generalized to
\cite{MST,GL,KPW}:
\begin{eqnarray}
\label{off}
F_2^A(x,Q^2) &=&
\int dy \int dp^2\ \varphi(y,p^2,Q^2)\ F_2^N(x^\prime,p^2,Q^2)\ ,
\end{eqnarray}
where $x^\prime=x/y$ and the function $\varphi(y,p^2,Q^2)$ depends on
the nuclear wave functions.
In the absence of $p^2$ dependence in $F_2^N$, the light-cone momentum
distribution $f(y,Q^2)$ in Eq.~(\ref{convolution}) would correspond to
the $p^2$ integral of $\varphi(y,p^2,Q^2)$.
In the approach of Ref.~\cite{GL}, the medium modified nucleon structure
function $F_2^N(x^\prime,p^2,Q^2)$ can be evaluated in terms of a
relativistic quark spectral function, $\rho_N$, as:
\begin{equation}
\label{off2}
F_2^N(x^\prime,p^2,Q^2) = \frac{{x^\prime}^2}{1-x^\prime} 
\sum_X \int_{k^2_{\rm min}} \frac{d k^2}{4 (2 \pi)^3}
        \rho_N(k^2(p),p_X^2)\ ,
\end{equation} 
where $\rho_N$ depends on the virtualities of the struck quark, $k^2$,
and spectator system, $p_X^2$, and the limit
$k_{\rm min} = k_{\rm min}(x^\prime,p^2,p_X^2)$ follows from the
positivity constraint on the struck quark's transverse momentum,
$k_\perp^2 \geq 0$.
The dependence of $k_{\rm min}$ on $p^2$ ($\neq M^2$) generates an
off-shell correction which grows with $A$ due to the $A$-dependence of
the virtuality $p^2$ of the bound nucleon.
This serves to enhance the EMC effect at large $x$ in comparison with
naive binding model calculations which do not take into account nucleon
off-shell effects \cite{CL}.
Assuming that the spectator quarks can be treated as a single system
with a variable mass $m_X^2$, the off-shell structure function in
Eq.~(\ref{off2}) can be related to the on-shell function by a
$p^2$-dependent rescaling of the argument $x^\prime$, namely \cite{GL}:
\begin{eqnarray}
\left. F_2^N(x^\prime)\right|_{p^2 \neq M^2}
&\longrightarrow&
\left. F_2^N(x^\prime(p^2) > x^\prime)\right|_{p^2=M^2}\ .
\end{eqnarray}
It is this (further) rescaling in $x$ that is responsible for the
larger effect at large $x$.

The effect of the off-shell correction on the ratio ${\cal R}$,
illustrated in Fig.~9, is a small ($\lesssim 1\%$) increase in the
ratio at $x \sim 0.6$.
Off-shell effects of this magnitude can be expected in models of the EMC
effect where the overall modification of the nuclear structure function
arises from a combination of conventional nuclear physics phenomena
associated with nuclear binding, and a small medium dependence of the
nucleon's intrinsic structure \cite{GST,MST,KPW,K2}.

Other models of the EMC effect, such as the color screening model for
the suppression of point-like configurations (PLC) in bound nucleons
\cite{FS85}, attribute most or all of the EMC effect to a medium
modification of the internal structure of the bound nucleo, and
consequently predict larger deviations of ${\cal R}$ from unity
\cite{SSS}.
However, recent $^4$He$(\vec e, e' \vec p)$ polarization transfer
experiments \cite{HE4} indicate that the
magnitude of the off-shell deformation is indeed rather small.
The measured ratio of transverse to longitudinal polarization of the
ejected protons in these experiments can be related to the medium
modification of the electric to magnetic elastic form factor ratio.
Using model independent relations derived from quark-hadron duality,
the medium modifications in the form factors were related to a
modification at large $x$ of the deep inelastic structure function of
the bound nucleon in Ref.~\cite{MTT}.
In $^4$He, for instance, the effect in the PLC suppression model was
found \cite{MTT} to be an order of magnitude larger than that allowed
by the data \cite{HE4}, and with a different sign for $x \agt 0.65$.
The results therefore place rather strong constraints on the size of
the medium modification of the structure of the nucleon, suggesting
little room for large off-shell corrections, and support a conventional
nuclear physics description of the $^3$He/$^3$H system as a reliable
starting point for nuclear structure function calculations.

% .......................................................................
\subsection{Nuclear Density Extrapolation Model}

The nuclear density model, which has proven successful for studying the
$A$-dependence of the EMC effect for heavy nuclei, stems from the 
empirical observation that for heavy nuclei the deviation from unity
in the EMC ratio $R(A)$ is assumed to scale with nuclear density
\cite{FS88}:
\begin{eqnarray}
\label{dens}
{ R(A_1)-1 \over R(A_2)-1 }
&=& { \rho(A_1) \over \rho(A_2) },
\end{eqnarray}
where $\rho(A) = 3A / (4\pi R_A^3)$ is the mean nuclear density
and $R_A^2 = (5/3) \langle r^2 \rangle_A$.
Whether the concept of density is physically meaningful for a few
body system such as a $^3{\rm He}$ nucleus is rather questionable
\cite{MABT}.
However, one can use the density extrapolation ansatz to investigate
the range of predictions for ${\cal R}$, and estimate the total
theoretical uncertainty.

{}From the empirical $A=3$ charge radii \cite{ATOM} one finds that
$\rho(^3{\rm H})/\rho(^3{\rm He}) \approx 140\%$, so that the EMC
effect in $^3$H is predicted to be 40\% bigger than in $^3$He.
However, as shown in Fig.~10, assuming that $R(^3{\rm He})$ can be
extrapolated from the measured EMC ratios for heavy nuclei such as
$^{56}$Fe, one still finds that ratio $|{\cal R}-1| < 2\%$ for all
$x \alt 0.85$.
The $x$-dependence predicted by density extrapolation method lies
approximately between that using the standard Faddeev and variational
techniques for $0.5 \lesssim x \lesssim 0.85$.

% .......................................................................
\subsection{Six-Quark Clusters}

While most of the medium modification of the nuclear structure function
at large $x$ can be described in terms of incoherent scattering from
bound nucleons, other effects involving explicit quark degrees of
freedom have been suggested as possible sources of EMC-type
modifications.
In particular, at short nucleon--nucleon separations the effects of
quark exchange could be more prominent.
Corrections to the impulse approximation arising from the exchange of
quarks between nucleons in $A=3$ nuclei were in fact discussed in
Ref.~\cite{HJ} (see also \cite{BKM}).
There the effect on the EMC ratio, for the isospin-averaged $A=3$
nucleus, was found to be comparable to that arising from binding.
However, the analysis \cite{HJ} did not allow for $NN$ correlations,
which are important at large momentum (and hence large $x$), so that
the overall EMC effect is likely to have been overestimated.

The effects of quarks which are not localized to single nucleons can
alternatively be parameterized in terms of multi-quark clusters, in
which six (or more) quarks form color singlets inside nuclei \cite{6Q}.
Six-quark configurations in the deuteron and other nuclei have been
studied in a variety of observables, including nuclear electromagnetic
form factors, $NN$ scattering, as well as the EMC effect.
To test the possible role of quark exchange on the ratio ${\cal R}$, we
consider the effect of six-quark clusters on $^3$He and $^3$H structure
functions (contributions from nine-quark clusters are presumably small
compared with those from six-quark states).
Although neither the normalization of the six-quark component of the
$A=3$ wave function, nor its momentum distribution, is known, one can
nevertheless estimate their potential importance by examining the
effect on ${\cal R}$ for a range of parameters.

Following Ref.~\cite{6Q}, contributions from scattering off quarks in a
six-quark cluster can be approximated by an effective six-quark structure
function, $F_2^{6q}(x_{6q})$, in the nucleus, where
$x_{6q} = Q^2/2M_{6q}\nu \approx x/2$.
If $P_{6q}$ is the probability of finding a six-quark cluster in
the nucleus, the net effect on the $^3$He (and similarly $^3$H)
structure function can be approximated by:
\begin{eqnarray}
F_2^{^3{\rm He}} &\longrightarrow&
(1-P_{6q}) F_2^{^3{\rm He}} + P_{6q} F_2^{6q}\ ,
\end{eqnarray}
where $F_2^{^3{\rm He}}$ is the incoherent nucleon contribution.
Taking a typical valence-like shape for $F_2^{6q}$, with the large-$x$
behavior constrained by hadron helicity counting rules,
$F_2^{6q} \sim (1-x_{6q})^9$, the effect on ${\cal R}$ is shown in
Fig.~11 for $P_{6q}=0\%$, 2\% and 4\%.
The overall effect is $\alt 1\%$ for all $x \alt 0.85$ even for the
largest six-quark probability considered.
For larger values of $P_{6q}$ the deviation from unity is in fact
even smaller, cancelling some of the effects associated with nucleon
off-shell dependence, for instance.
We have also considered other six-quark structure functions, and while
there is some sensitivity to the exact shape of $F_2^{6q}$, the
$\sim 1\%$ effect on ${\cal R}$ appears to be an approximate upper
limit for all $x$.

%%%%%%%%%%%%%%%%%%%%%%%%%%%%%%%%%%%%%%%%%%%%%%%%%%%%%%%%%%%%%%%%%%%%%%%%%%
\section{Experimental Considerations}

Measurements of the nucleon structure functions have been performed 
at several accelerator laboratories over the past 35 years.  The highest-$x$
measurements using proton and deuteron targets were part of the historic
Stanford SLAC-MIT experiments of the late 1960's and early 1970's
\cite{nobel}. 
The natural place to continue studies of the nucleon and nuclear
structure functions at high $x$ and moderate $Q^2$ is Jefferson
Lab (JLab) with its high intensity electron accelerator and large acceptance
spectrometer facilities.
The proposed energy upgrade \cite{12GEV} of the Continuous Electron Beam
Accelerator of JLab will offer a unique opportunity to 
perform electron deep inelastic scattering studies off the $A=3$ system,
as has been recently proposed \cite{phily,pac18}. 
The proposal calls for precise measurements of the $^3$He and $^3$H
inelastic cross sections, under identical conditions, using an 11~GeV
upgraded electron beam of JLab and the Hall A Facility of JLab.
The inelastic electron--nucleus cross section is given in terms of the
unpolarized structure functions $F_1$ and $F_2$ by:
\begin{equation}
\label{bla1}
{
{\sigma} \equiv
 { {d^2\sigma} \over {d\Omega dE'} } (E_\circ,E',\theta) =
 { {4 \alpha^2 (E')^2} \over {Q^4} }  \cos^2(\theta/2)
\left[ { F_2^A(\nu,Q^2) \over \nu }
     + { 2 F_1^A(\nu,Q^2) \over {M_A} } \tan^2(\theta/2) \right],
}
\end{equation}
where $\alpha$ is the fine structure constant, $E_\circ$ is the incident
electron energy, $E'$ and $\theta$ are the scattered electron energy
and angle, $\nu=E_\circ-E'$ is the energy transfer, and $M_A$ is the nuclear
mass.

The structure functions $F_1^A$ and $F_2^A$ are connected through the
ratio $R^A = {\sigma_L^A / \sigma_T^A}$ by:
\begin{equation}
\label{dis1}
{
F_1^A = { {F_2^A (1 + Q^2/\nu^2)} \over {2 x (1+R^A)} }
}\ ,
\end{equation}
where $\sigma_L^A$ and $\sigma_T^A$ are the nuclear virtual
photoabsorption cross sections for longitudinally and transversely 
polarized photons.
The ratio $R^A$ has been measured to be independent of the mass number
$A$ in precise SLAC and CERN measurements using hydrogen, deuterium, 
iron and other nuclei (for a compilation of data see Ref.~\cite{GST}).

By performing the tritium and helium measurements, under identical
conditions using the same incident beam and scattered electron detection 
system configurations (same $E_\circ$, $E'$ and $\theta$), and assuming that 
the ratio $R^A$ is the same for both nuclei, the ratio of the
inelastic cross sections for the two nuclei provides a direct
measurement of the ratio of the $F_2$ structure functions:
\begin{equation}
\label{dis2}
{
{ {\sigma^{\rm ^3{\rm H }}(E_\circ,E',\theta)} \over 
  {\sigma^{\rm ^3{\rm He}}(E_\circ,E',\theta)} } =
{ F_2^{^3{\rm H }}(\nu,Q^2) \over   
  F_2^{^3{\rm He}}(\nu,Q^2) }
}\ .
\end{equation}

The key issue for this experiment will be the availability of a high 
density tritium target planned for the Hall A Facility of JLab \cite{rans}. 
Tritium targets have been used in the past to measure the elastic form 
factors of $^3$H at Saclay \cite{SACLAY} and MIT-Bates \cite{BATES}.
The Saclay target contained liquid $^3$H at 22~K and was able to tolerate 
beam currents up to 10~$\mu$A with very well understood beam-induced 
density changes.
The nominal tritium density of 0.271 g/cm$^3$ at the operating conditions
of this target was known, from actual density
measurements, to $\pm 0.5 \%$.
The MIT-Bates target contained $^3$H gas at 45~K and 15~atm,
and was able to tolerate beam currents up to 25~$\mu$A with small measurable 
beam-induced density changes. 
The tritium density, under these operating conditions, was determined to 
be 0.218~mg/cm$^2$ with $\pm2\%$ uncertainty, using the Virial formalism 
for hydrogen.

Given a high density tritium target, an entire program of elastic, 
quasi-elastic and inelastic measurements will be possible at JLab.
This program can be better accomplished by building a target similar to 
the one used at MIT-Bates (the cooling mechanism of a target similar to the
Saclay one would prevent coincidence measurements ).
The tritium density can be better determined from comparison of the 
elastic cross section measured with the 45~K/15~atm cell and a cell 
filled up with tritium at higher temperatures (ideal gas of known 
density).
Two more cells will also be necessary for the $^3$He measurements.

The large solid angle and the wide kinematical coverage of the
proposed Medium Acceptance Device (MAD) Hall A
spectrometer~\cite{MAD} will facilitate precise inelastic
cross section measurements (statistical errors of $\le \pm 0.25 \%$)
in a large-$x$ range as well as valuable systematics checks
using reasonably short amounts of beam time.
An important systematic check would be the confirmation that the ratio 
$R$ is the same for $^3$H and $^3$He.
The performance of the above spectrometer is expected to be comparable,
if not better, to that of the SLAC 8~GeV/c spectrometer~\cite{bodek} that
has provided precise measurements for absolute inelastic cross sections,
inelastic cross section ratios, and differences in $R$ for several
nuclei~\cite{GOMEZ,DA94,TAO}.
The overall systematic errors for these measurements have been typically 
$\pm2\%$, $\pm0.5\%$ and $\pm 0.01\%$, respectively.
Since the objective of the experiment is the measurement of cross section
ratios rather than absolute cross sections, many of the experimental
errors that plague absolute measurements will cancel out.
The experimental uncertainties on the ratio of cross sections should
be similar to those achieved by SLAC experiments E139~\cite{GOMEZ}
and E140~\cite{DA94,TAO}, which were typically around $\pm0.5\%$.

Deep inelastic scattering with an upgraded 11~GeV JLab electron beam can
provide measurements for the $^3$H and $^3$He $F_2$ structure functions
in the $x$ range from 0.10 to 0.82.
The electron scattering angle will range from 12$^\circ$ to 47$^\circ$
and the electron scattered energy from 1.0 to 6.0~GeV.
It is assumed that the MAD spectrometer system will be instrumented with
a threshold gas $\check{\rm C}$erenkov counter and a segmented lead-glass
calorimeter,
which will provide discrimination between scattered electrons and an
associated hadronic (mostly pion) background.
The above two-counter combination has provided in the past a pion
rejection factor of at least 10,000 to 1~\cite{DA94} that has allowed
inelastic cross section measurements with negligible pion
contamination 
for cases where the ratio of pion background to electron signal 
($\pi / e$) was as large as 300.
The expected $\pi/e$ ratio for this experiment has been estimated using
SLAC data from measurements of photon-nucleon cross sections~\cite{WISER} 
and is
less than 300.
The estimated $\pi / e$ ratios are given in Table~I along with the
kinematical parameters for the proposed ``core'' set of measurements
of the ratio  $F_2^{^3\rm H}/F_2^{^3\rm He}$ up to $x=0.82$.

The estimated inelastic cross sections, counting rates and the beam
time required
for the above measurements are given in Table~II, assuming $^3$H and
$^3$He luminosities of $\sim 5 \times 10^{37}$ cm$^{-2}$ s$^{-1}$.
The rates have been estimated under the assumption that
$\sigma^{^3{\rm He}} \simeq \sigma_d + \sigma_p$ and
$\sigma^{^3{\rm H}}  \simeq 2 \sigma_d - \sigma_p$, using values for
the proton ($\sigma_p$) and deuteron ($\sigma_d$) inelastic cross sections
and for the ratio $R$ from the SLAC ``global'' analysis~\cite{WHITLOW}
of all available SLAC data.
The rates are based on the MAD design specifications and include an
approximation of radiative effects.
It is evident from the listed rates that the proposed experiment will
be able to provide very high statistics data and perform necessary
systematic studies in a timely fashion.

The 11~GeV beam and the momentum and angular range of MAD will allow 
measurements of $R$ in the same $x$ range as in the SLAC experiments
by means of a Rosenbluth separation versus
$\epsilon = \left[ 1 + 2 (1+\nu^2/Q^2) \tan^2(\theta/2) \right]^{-1}$
(the degree of the longitudinal polarization of the virtual photon 
mediating the scattering).
The $R$ measurements will be limited by inherent systematics
uncertainties rather than statistical uncertainties as in the SLAC case.
It is estimated that the $R$ measurements will require an amount of beam 
time comparable to the one required for the core set of measurements 
listed in Table~II.

The $F_2^{^3\rm H}/F_2^{^3\rm He}$ ratio is expected to be dominated by
experimental uncertainties that do not cancel in the inelastic cross section 
ratio of $^3\rm H$ to $^3\rm He$ and by the theoretical uncertainty in the 
calculation of the ratio $\cal R$.
Assuming that the target densities can be known to the $\pm0.5\%$ level 
and that the relative difference in the $^3$H and $^3$He radiative 
corrections would be $\pm0.5\%$ as in Refs.~\cite{GOMEZ,DA94}, the total
experimental error in the the inelastic cross section ratio of $^3\rm H$ to 
$^3\rm He$ should be $\sim\pm1.0\%$.
Such an error is comparable to a realistic maximum theoretical
uncertainty ($\sim\pm1\%$ in the vicinity of $x$~=~0.8) in the
calculation of the ratio $\cal R$.

The quality of the expected $F_2^n/F_2^p$ extracted values is shown in
Fig.~12.
The two sets of data in this Figure represent the extreme possible
values for the ratio  $F_2^n/F_2^p$ (see Fig.~1) and are indicative of 
the present 
uncertainties in the nuclear corrections in the extraction of $F_2^n/F_2^p$ 
from proton and deuterium inelastic scattering data.
The shaded band represents the projected uncertainty 
($\pm$ one standard deviation
error band) of the proposed JLab measurement.
The band assumes a $\pm1\%$ overall systematic experimental error 
in the measurement of the $\sigma^{^3\rm H}/\sigma^{^3\rm He}$ ratio and a 
theoretical uncertainty in $\cal R$ that increases linearly from 0\%
at $x=0$ to $\pm1\%$ at $x=0.82$.
The central value of the projected JLab band has been arbitrarily
chosen, for this
comparison purpose, to follow the trend obtained in the relativistic
analysis of nuclear binding and Fermi motion of Ref.~\cite{MT} (see Fig.~1).
It is evident that the proposed measurement will be able to
unquestionably distinguish between the present competing extractions of
the $F_2^n/F_2^p$ ratio from proton and deuterium inelastic measurements, and 
determine its value with an unprecedented precision in an almost 
model-independent way.

A secondary goal of this proposed experiment would be the precise
determination of the EMC effect in $^3$H and $^3$He.
At the present time, the available SLAC and CERN data allow for two 
equally compatible parametrizations~\cite{GOMEZ} of the EMC effect,
within the achieved experimental uncertainties.
In the first parametrization the EMC effect is parametrized versus the mass number $A$,
and in the second one versus the nuclear density, $\rho$.
While the two parametrizations are indistinguishable for heavy nuclei,
they predict quite distinct patterns for $A=3$.
The expected precision ($\pm 1\%$) of this experiment for the
$F_2^{^3{\rm H}}/F_2^{^3{\rm He}}$ ratio should easily allow to 
distinguish between the two competing parametrizations.
The proposed measurements should bring a closure to the EMC effect
parametrization issue and provide crucial input for a more complete
explanation of the origin of the EMC effect.

%%%%%%%%%%%%%%%%%%%%%%%%%%%%%%%%%%%%%%%%%%%%%%%%%%%%%%%%%%%%%%%%%%%%%%%%%%
\section{Conclusion}

We have presented a comprehensive analysis of deep inelastic scattering
from $^3$He and $^3$H nuclei, focusing in particular on the extraction
of the free neutron structure function at large $x$.
We have demonstrated the effectiveness of using the mirror symmetry of
$A=3$ nuclei to extract the ratio of the neutron to proton structure
functions, $F_2^n/F_2^p$, free of nuclear effects to $\alt 1$--2\% for
all $x \alt 0.8$.
This is comparable with the expected experimental errors for the
simultaneous measurement of $^3$He and $^3$H DIS cross sections at an
energy-upgraded Jefferson Lab, for instance.

The major theoretical uncertainty involved in the extraction is that
associated with the nuclear wave functions of $^3$He and $^3$H.
We have examined two independent methods of calculating the nuclear
spectral function, namely by solving the Faddeev equations, and using
a variational approach, for a range of two-body interactions.
The resulting structure function ratios have been studied as a function
of $x$ and $Q^2$ for various input nucleon structure function
parameterizations.
By utilizing an iterative procedure, the dependence of the extracted
$F_2^n$ on input parameterizations can be effectively removed altogether.
We find this procedure converges quite rapidly, requiring only $\sim 3$
iterations.

We have also considered explicit charge symmetry breaking effects in
the nuclear wave functions, and effects associated with the medium
modification of the bound nucleon structure functions, as well as
corrections to the impulse approximation arising from non-nucleonic
degrees of freedom such as six-quark clusters.
In all cases consistent with existing nuclear phenomenology we find
that the nuclear effects in the ratio ${\cal R}$ of $^3$He to $^3$H
EMC ratios cancel to within 1--2\% at the relevant kinematics, making
this an extremely robust method with which to extract the free neutron
structure function, and thus settle a ``text-book'' issue which has
eluded a definitive resolution for nearly three decades.

Once the $F_2^n/F_2^p$ ratio is determined, one can combine the free
proton and deuteron data to obtain the size of the EMC effect in the
deuteron, which remains a source of controversy, via:
\begin{eqnarray}
R(d)
&=& { F_2^d \over
      F_2^p \left( 1 + \left. F_2^n/F_2^p\right|_{\rm extr} \right) }\ ,
\end{eqnarray}
where $\left. F_2^n/F_2^p\right|_{\rm extr}$ is the neutron to proton
ratio extracted from Eq.~(\ref{np}).

While the ratio ${\cal R}$ is not very sensitive to nuclear dynamics
in the $A=3$ system, measurement of the absolute $^3$He to $^3$H cross
sections will, on the other hand, enable one to discriminate between
different models.
In particular, it will allow the completion of the empirical study of
nuclear effects in deep inelastic scattering over the full range of
mass numbers.

%%%%%%%%%%%%%%%%%%%%%%%%%%%%%%%%%%%%%%%%%%%%%%%%%%%%%%%%%%%%%%%%%%%%%%%%%
\acknowledgements

This work was supported by the Australian Research Council, the U.S.
Department of Energy contract \mbox{DE-AC05-84ER40150}, under which the
Southeastern Universities Research Association (SURA) operates the
Thomas Jefferson National Accelerator Facility (Jefferson Lab), the U.S.
Department of Energy contract \mbox{DE-FG02-01ER41200}, and the
U.S. National Science Foundation grant PHY-0072384.

%%%%%%%%%%%%%%%%%%%%%%%%%%%%%%%%%%%%%%%%%%%%%%%%%%%%%%%%%%%%%%%%%%%%%%%%%

%%%%%%%%%%%%%%%%%%%%%%%%%%%%%%%%%%%%%%%%%%%%%%%%%%%%%%%%%%%%%%%%%%%%%%%%%%%%
\newpage

\begin{table}[h]
\begin{center}
% \begin{tabular}{lllllc}
\begin{tabular}{lccccc}
%\multicolumn{6}{c}{\bf Helium/Tritium DIS Kinematics with ${\mathbf E}$~=~11~GeV} \\ \\  
\multicolumn{6}{c} {}\\ \hline \\
$x$     &  $W^2$         &  $Q^2$         & $E^\prime$ & $\theta$ &  $\pi/e$  \\ 
      &  [(GeV/c)$^2]$ &  [(GeV/c)$^2$] & (GeV)      &\ \ (deg)\ \ & \\ \\ \hline \\
0.82  &  4.0           &     13.8   &  2.00      &  46.6    &    52     \\  
0.77  &  4.7           &     12.9   &  2.10      &  43.8    &    43     \\
0.72  &  5.5           &     11.9   &  2.20      &  41.0    &    36     \\
0.67  &  6.2           &     10.9   &  2.35      &  37.8    &    27     \\
0.62  &  6.9           &      9.8   &  2.55      &  34.4    &    19     \\
0.57  &  7.6           &      8.9   &  2.65      &  32.1    &    19     \\
0.52  &  8.3           &      8.1   &  2.75      &  29.9    &    18     \\
0.47  &  9.0           &      7.2   &  2.85      &  27.7    &    19     \\
0.42  &  9.6           &      6.3   &  3.00      &  25.2    &    18     \\
0.37  &  10.2          &      5.5   &  3.10      &  23.1    &    19     \\
0.32  &  10.7          &      4.6   &  3.30      &  20.6    &    18     \\
0.27  &  11.2          &      3.8   &  3.50      &  18.1    &    18     \\
0.22  &  11.6          &      3.0   &  3.65      &  15.8    &    19       
 \\ \\ \hline              
 
\end{tabular}
\end{center}
\caption{Kinematics of the proposed JLab experiment~\cite{phily,pac18} 
on the measurement of the $F_2^n/F_2^p$ ratio using $^3$H and $^3$He
targets for an
incident electron energy of 11 GeV (see text).  The parameter $W^2$ is the
invariant mass of the final hadronic state.  The last column is the
estimated ratio of the pion background to the scattered electron signal.}
\end{table}

\begin{table}[h]
\begin{center}
% \begin{tabular}{llllllc}
\begin{tabular}{lcccccc}
%\multicolumn{7}{c}{\bf Cross Sections, Counting Rates and Beam Times} \\ \\  
\multicolumn{7}{c} {}\\ \hline \\
$x$       & $\sigma^{3\rm He}$ & $\sigma^{3\rm H}$  & $^3$He Rate & $^3$H Rate   &  $^3$He Time &  $^3$H Time  \\  
      &  (nb/sr/GeV) & (nb/sr/GeV) & (Events/h)        & (Events/h)
      & (h)  & (h)    \\ \\ \hline \\
0.82  &  0.0146      & 0.0117      & 1.55 $\cdot$ $10^4$ & 1.25 $\cdot$ $10^4$
      & 10.3 & 12.8  \\  
0.77  &  0.0308      & 0.0240      & 3.55 $\cdot$ $10^4$ & 2.77 $\cdot$ $10^4$
      & 4.5  & 5.8   \\
0.72  &  0.0639      & 0.0491      & 8.01 $\cdot$ $10^4$ & 6.16 $\cdot$ $10^4$
      & 2.0  & 2.6   \\
0.67  &  0.130       & 0.0996      & 1.80 $\cdot$ $10^5$ & 1.38 $\cdot$ $10^5$
      & 0.9  & 1.2   \\
0.62  &  0.261       & 0.202       & 4.02 $\cdot$ $10^5$ & 3.12 $\cdot$ $10^5$
      & 0.5  & 0.5   \\
0.57  &  0.463       & 0.364       & 7.76 $\cdot$ $10^5$ & 6.10 $\cdot$ $10^5$
      & 0.5  & 0.5   \\
0.52  &  0.801       & 0.639       & 1.43 $\cdot$ $10^6$ & 1.14 $\cdot$ $10^6$
      & 0.5  & 0.5   \\
0.47  &  1.35        & 1.10        & 2.51 $\cdot$ $10^6$ & 2.04 $\cdot$ $10^6$
      & 0.5  & 0.5   \\
0.42  &  2.35        & 1.95        & 4.58 $\cdot$ $10^6$ & 3.80 $\cdot$ $10^6$
      & 0.5  & 0.5   \\
0.37  &  3.89        & 3.30        & 7.84 $\cdot$ $10^6$ & 6.65 $\cdot$ $10^6$
      & 0.5  & 0.5   \\
0.32  &  7.00        & 6.07        & 1.50 $\cdot$ $10^7$ & 1.30 $\cdot$ $10^7$
      & 0.5  & 0.5   \\
0.27  &  12.8        & 11.3        & 2.91 $\cdot$ $10^7$ & 2.58 $\cdot$ $10^7$
      & 0.5  & 0.5   \\
0.22  &  23.3        & 21.1        & 5.53 $\cdot$ $10^7$ & 5.01 $\cdot$ $10^7$ 
      & 0.5 & 0.5
 \\ \\ \hline              
 
\end{tabular}
\end{center}
\caption{Estimated values of the $^3$He and $^3$H inelastic cross
  sections for the kinematics of Table~I, expected scattered
  electron counting rates using JLab Hall A planned facilities (see
  text), and required amounts of beam time for $\pm$0.25\% cross section
  statistical uncertainties.}
\end{table}

%%%%%%%%%%%%%%%%%%%%%%%%%%%%%%%%%%%%%%%%%%%%%%%%%%%%%%%%%%%%%%%%%%%%%%%%%
\begin{figure}  % FIG 1
\begin{center}
\epsfig{figure=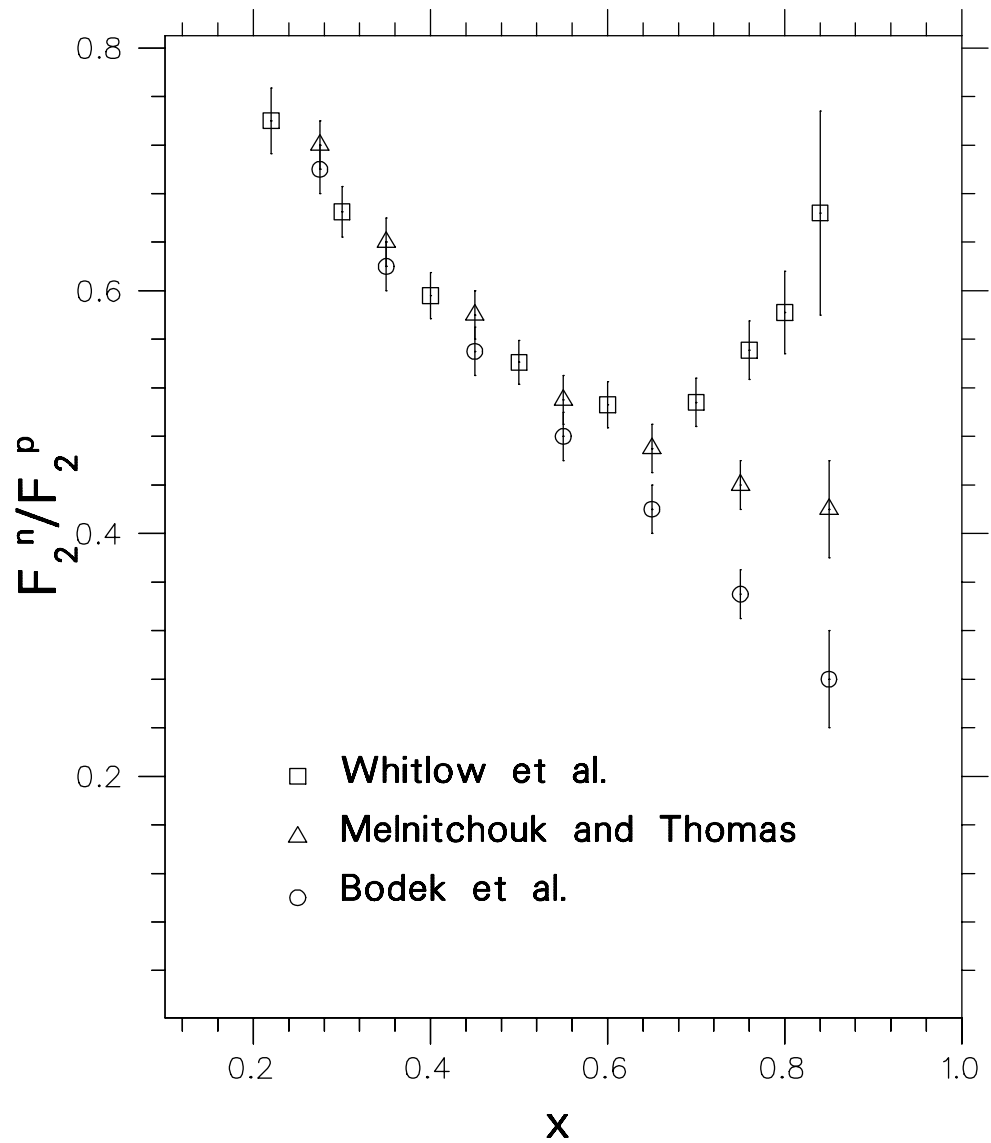,height=15cm}
\vspace*{1cm}
\caption{Neutron to proton ratio, extracted from inclusive proton and
        deuteron inelastic data, correcting for the effects of
        Fermi motion and nuclear binding (Melnitchouk \& Thomas
        \protect\cite{MT}), Fermi motion only (Bodek {\em et al.}
        \protect\cite{bodek}), and using the density extrapolation
        model (Whitlow {\em et al.} \protect\cite{WHITLOW}).}
\end{center}
\end{figure}

\begin{figure}  % FIG 2
\begin{center}
\epsfig{figure=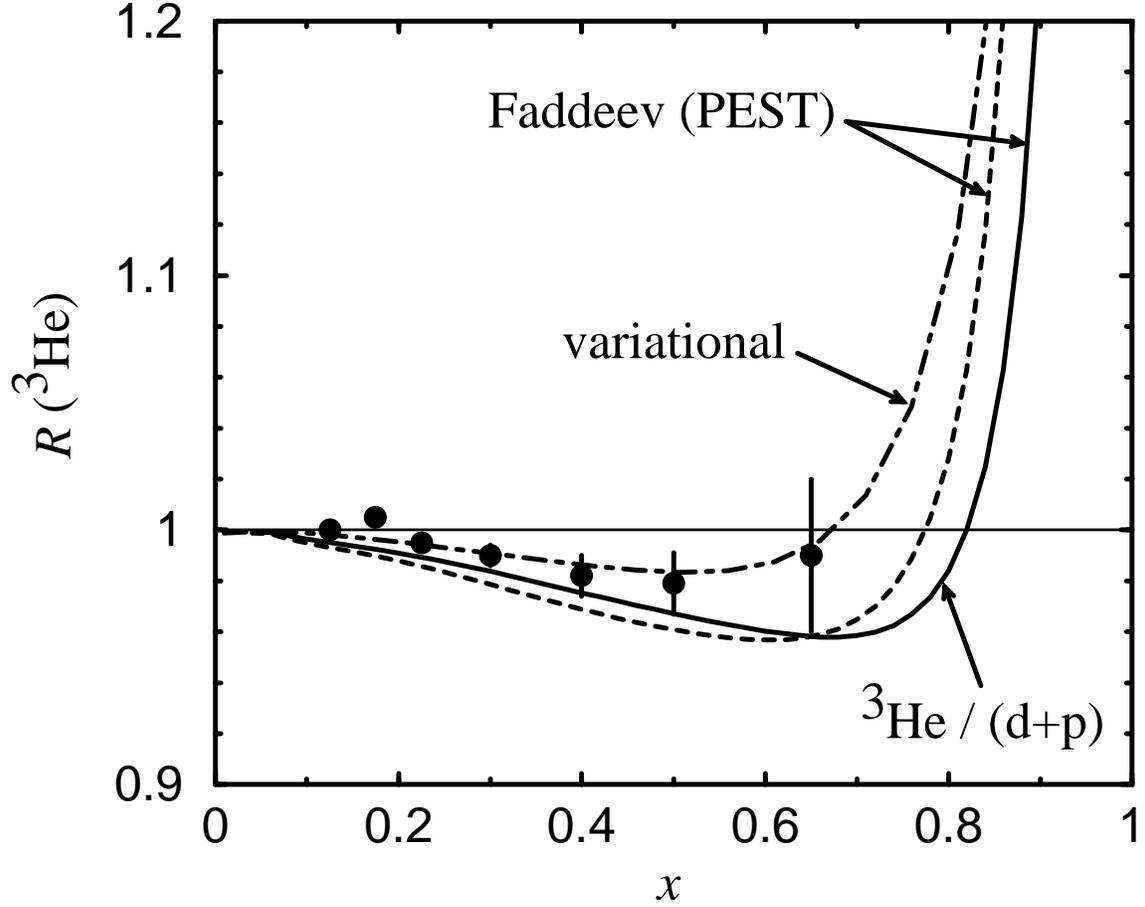,height=12cm}
\vspace*{1cm}
\caption{Nuclear EMC ratio in $^3$He using the Faddeev (with the PEST
        potential) and variational (RSC) wave functions, compared with
        HERMES data \protect\cite{HERMES} for
        $\sigma(^3{\rm He})/(\sigma(d)+\sigma(p))$.
        The solid curve corresponds to $F_2^{^3{\rm He}}/(F_2^d + F_2^p)$,
        while the dashed and dot-dashed assume no EMC effect in the
        deuteron.}
\end{center}
\end{figure}

\begin{figure}  % FIG 3
\begin{center}
\epsfig{figure=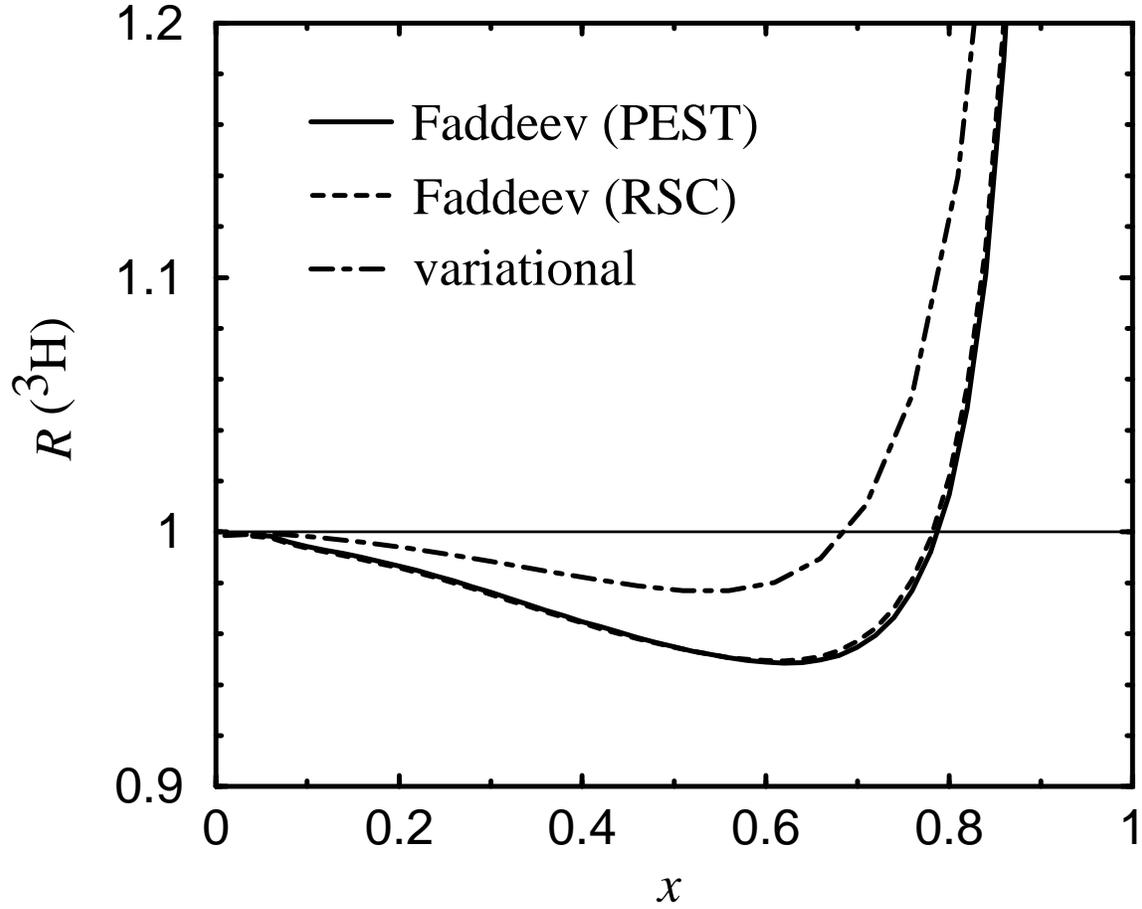,height=12cm}
\vspace*{1cm}
\caption{Nuclear EMC ratio in $^3$H using the Faddeev (with PEST and RSC
        potentials) and variational (RSC) wave functions.}
\end{center}
\end{figure}

\begin{figure}  % FIG 4
\begin{center}
\epsfig{figure=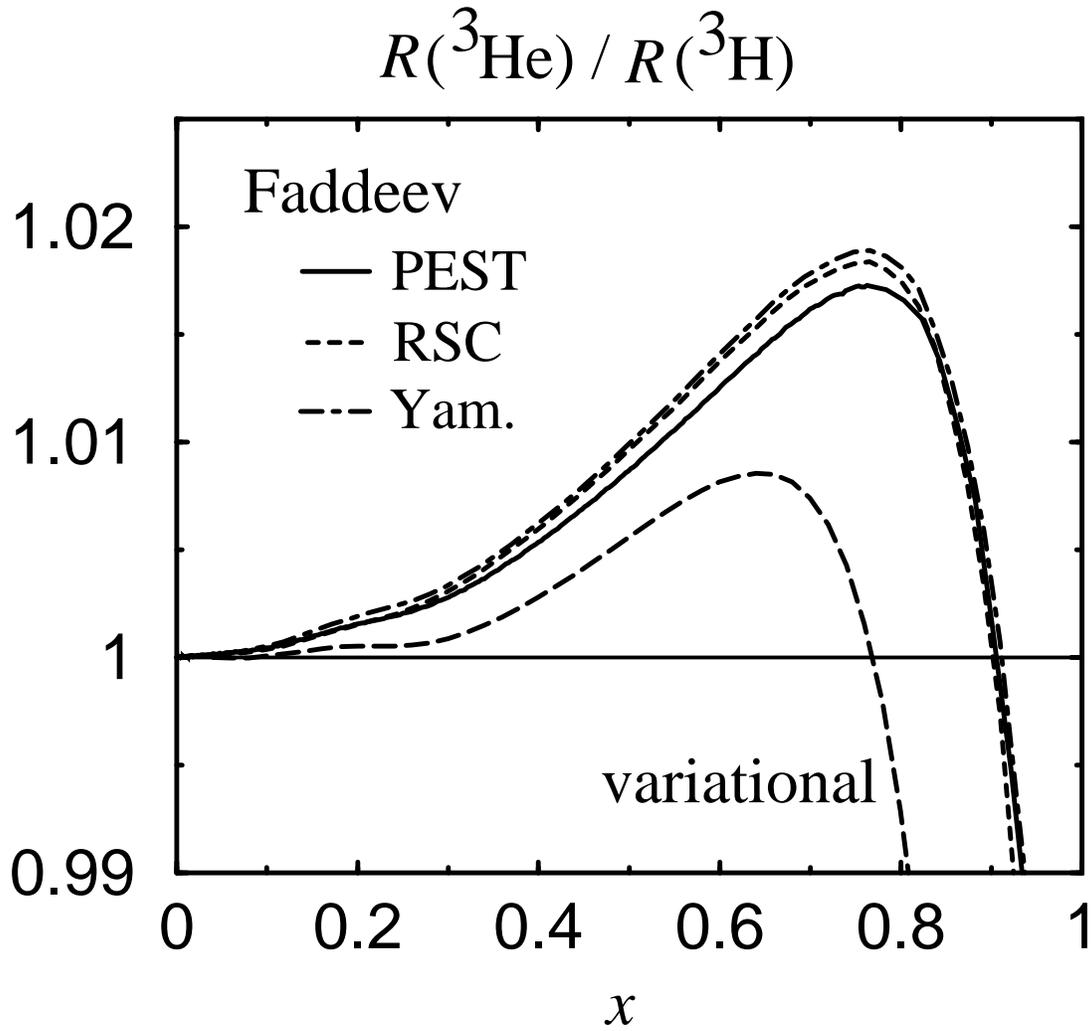,height=14cm}
\vspace*{1cm}
\caption{Ratio ${\cal R}$ of nuclear EMC ratios for $^3$He and $^3$H
        nuclei, with the nucleon momentum distribution calculated from
        the Faddeev (PEST, RSC, Yamaguchi) and variational (RSC)
        wave functions.}
\end{center}
\end{figure}

\begin{figure}  % FIG 5
\begin{center}
\epsfig{figure=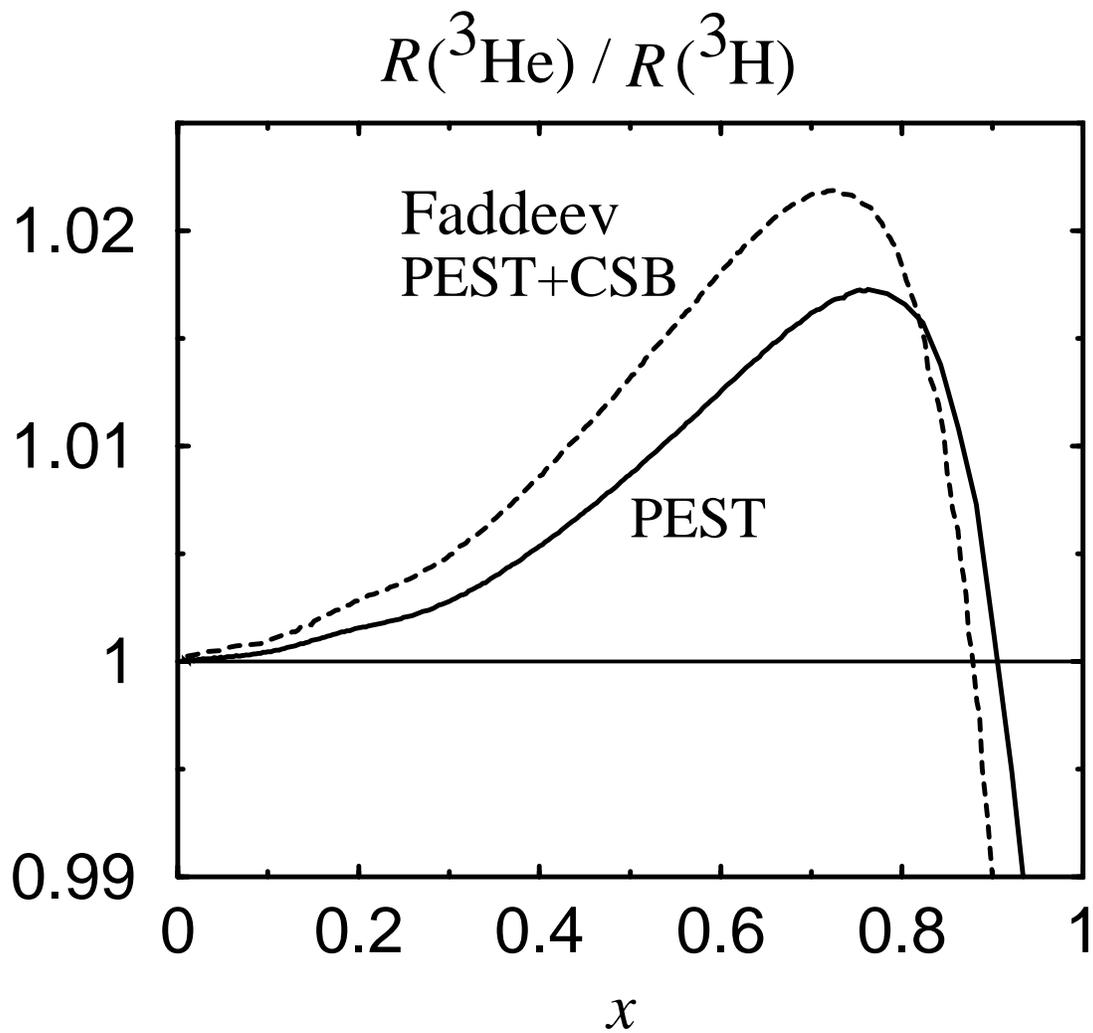,height=14cm}
\vspace*{1cm}
\caption{Ratio of nuclear EMC ratios for $^3$He and $^3$H for the Faddeev
        (PEST) wave function, with (dashed) and without (solid) charge
        symmetry breaking (CSB) effects.}
\end{center}
\end{figure}

\begin{figure}  % FIG 6
\begin{center}
\epsfig{figure=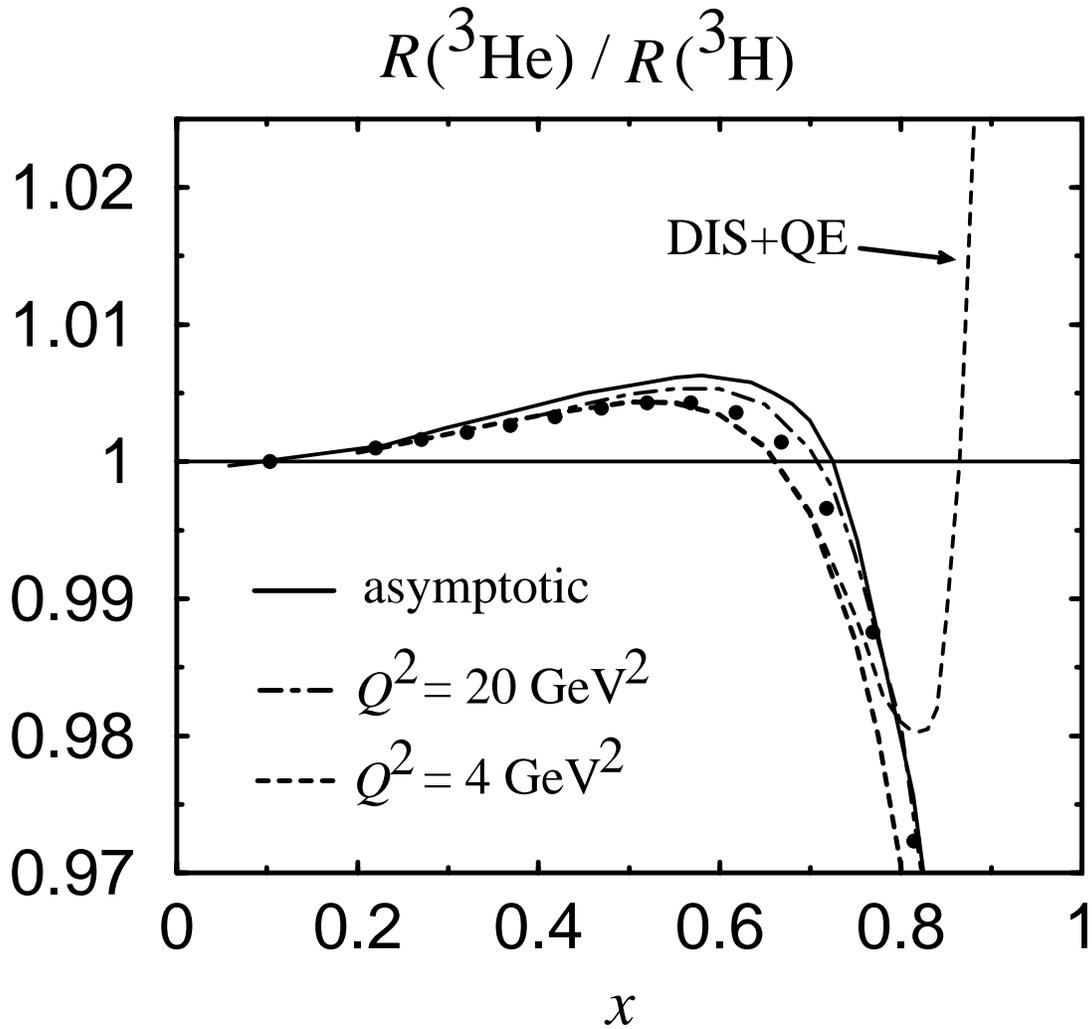,height=14cm}
\vspace*{1cm}
\caption{$Q^2$ dependence of the ratio ${\cal R}$ of $^3$He and $^3$H
        nuclear EMC ratios, for the variational wave functions with
        the RSC potential.
        Results using the full $Q^2$ dependence in
        Eq.~(\protect\ref{F2Afull}) at $Q^2=4$~GeV$^2$ (dashed) and
        $Q^2=20$~GeV$^2$ (dot-dashed) are compared with the asymptotic
        prediction (solid), and at varying $Q^2$ (filled circles)
        ranging from $Q^2=3$~GeV$^2$ for the lowest $x$ bin
        to $Q^2=14$~GeV$^2$ at the highest.
        The effect of the quasi-elastic contribution (DIS+QE)
        at $Q^2=4$~GeV$^2$ is also indicated.}
\end{center}
\end{figure}

\begin{figure}  % FIG 7
\begin{center}
\epsfig{figure=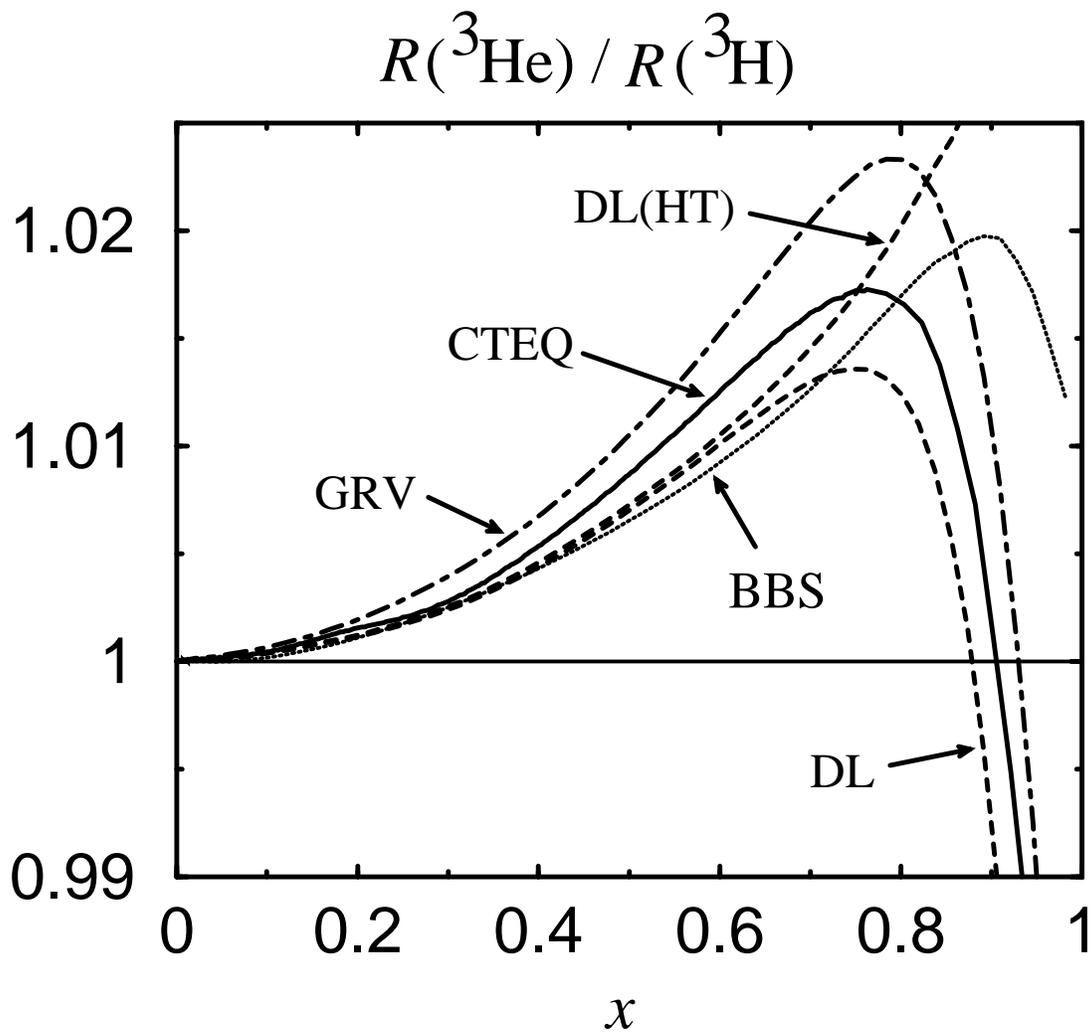,height=14cm}
\vspace*{1cm}
\caption{Ratio of nuclear EMC ratios for $^3$He and $^3$H with the
        the Faddeev (PEST) wave functions, for various nucleon
        structure function parameterizations:
        CTEQ \cite{CTEQ}, GRV \cite{GRV}, BBS \cite{BBS}, and
        Donnachie-Landshoff (DL) \cite{DOLA} with leading twist only,
        and with higher twist (HT) correction (dot-dashed).}
\end{center}
\end{figure}

\begin{figure}  % FIG 8
\begin{center}
\hspace*{-1cm}
\epsfig{figure=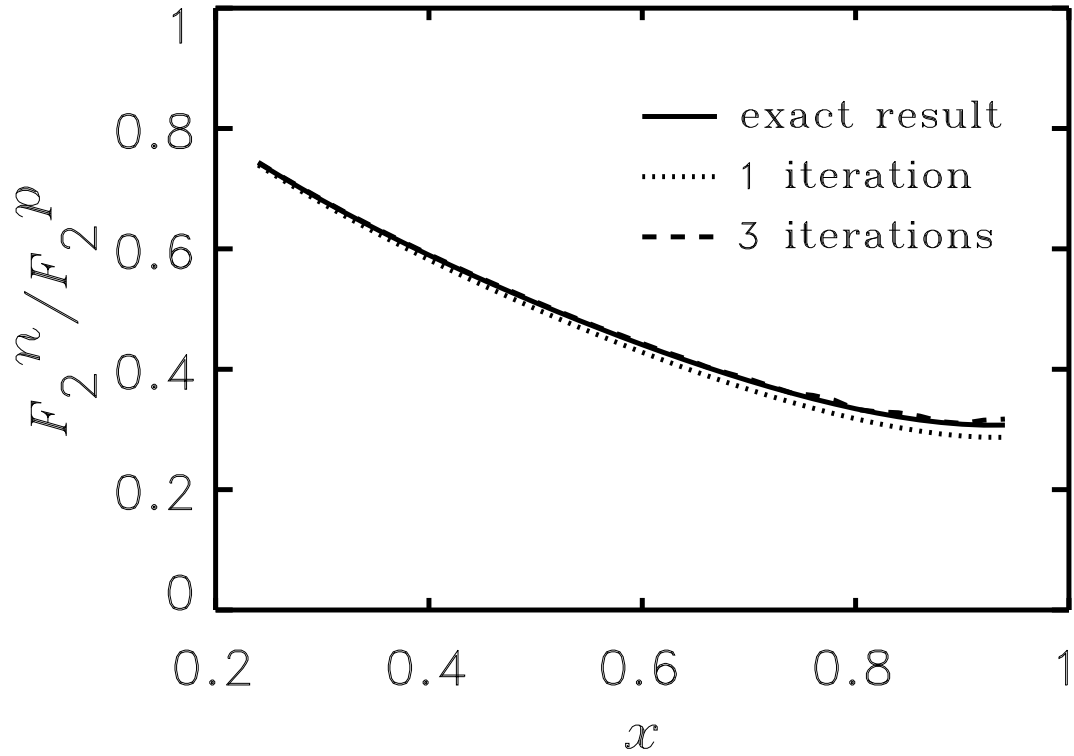,height=12cm}
\vspace*{1cm}
\caption{Neutron to proton structure function ratio extracted from the
        $F_2^{^3{\rm He}}/F_2^{^3{\rm H}}$ ratio via the iteration
        procedure.
        The input is $F_2^n/F_2^p=1$, and the ratio after $\sim 3$
        iterations is indistinguishable from the exact result.}
\end{center}
\end{figure}

\begin{figure}  % FIG 9
\begin{center}
\epsfig{figure=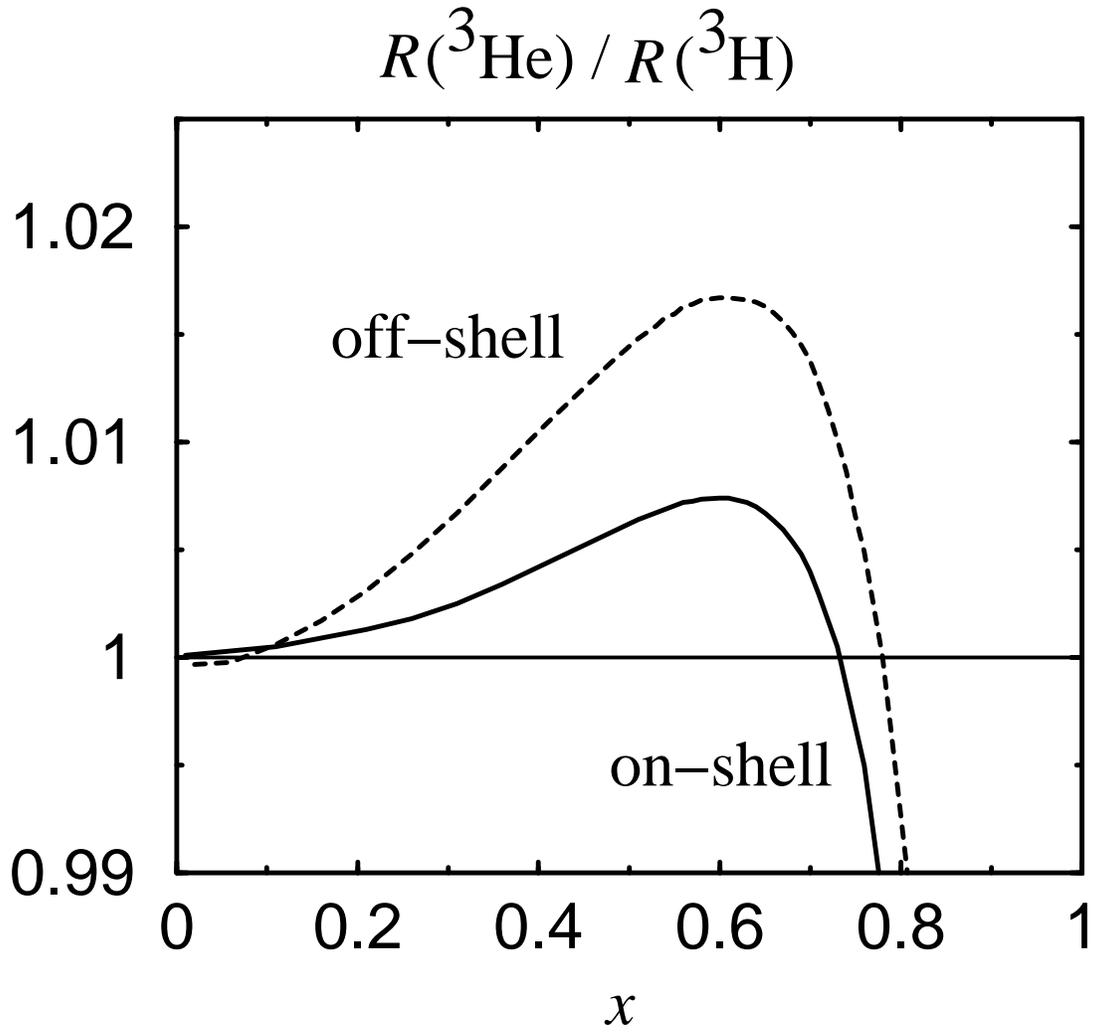,height=14cm}
\vspace*{1cm}
\caption{Ratio ${\cal R}$ of nuclear EMC ratios for $^3$He and $^3$H
        nuclei, with (dashed) and without (solid) nucleon off-shell
        corrections \protect\cite{GL} (see text), for the variational
        (RSC) wave function.}
\end{center}
\end{figure}

\begin{figure}  % FIG 10
\begin{center}
\epsfig{figure=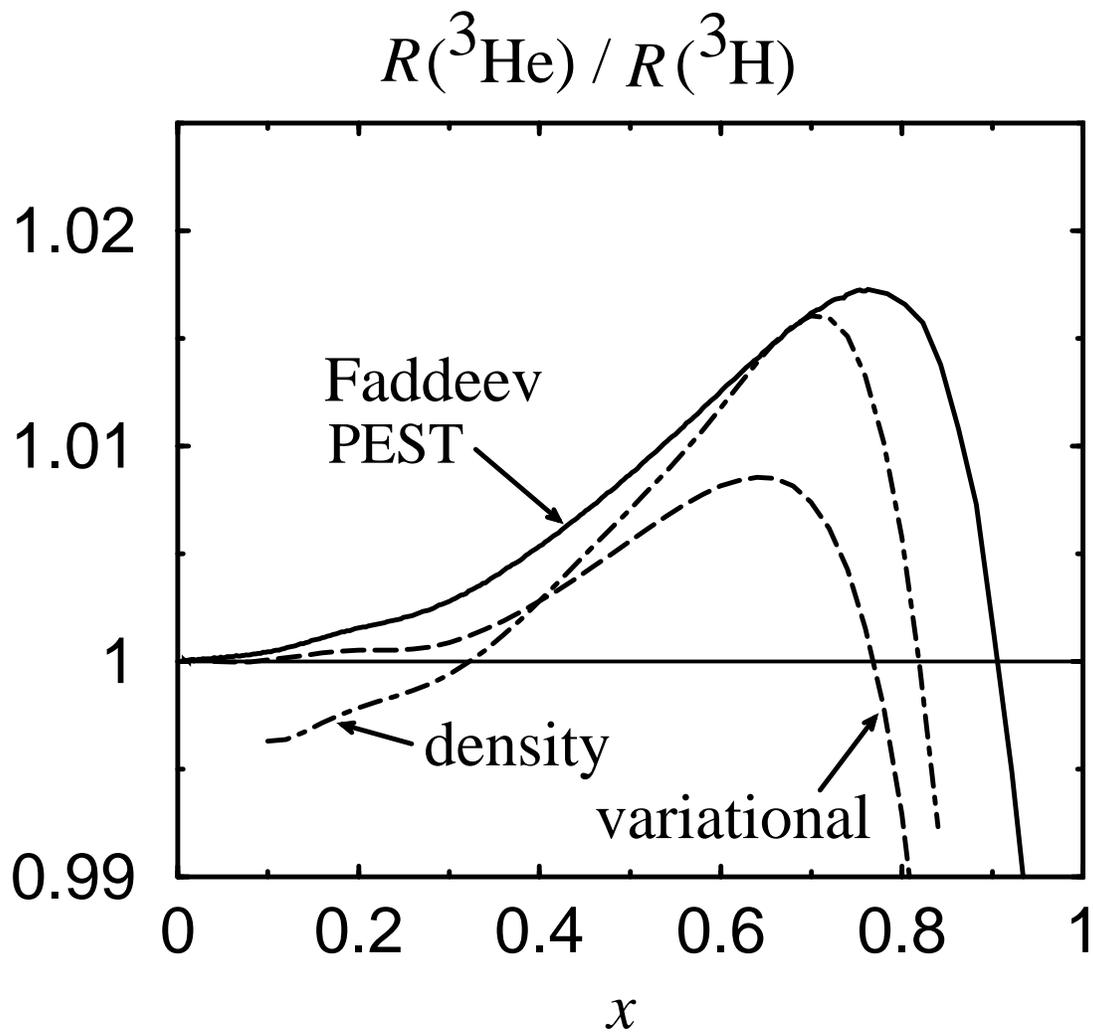,height=14cm}
\vspace*{1cm}
\caption{Ratio of nuclear EMC ratios for $^3$He and $^3$H for the
        density extrapolation model, compared with the standard
        Faddeev (PEST) and variational (RSC) wave functions.}
\end{center}
\end{figure}

\begin{figure}  % FIG 11
\begin{center}
\epsfig{figure=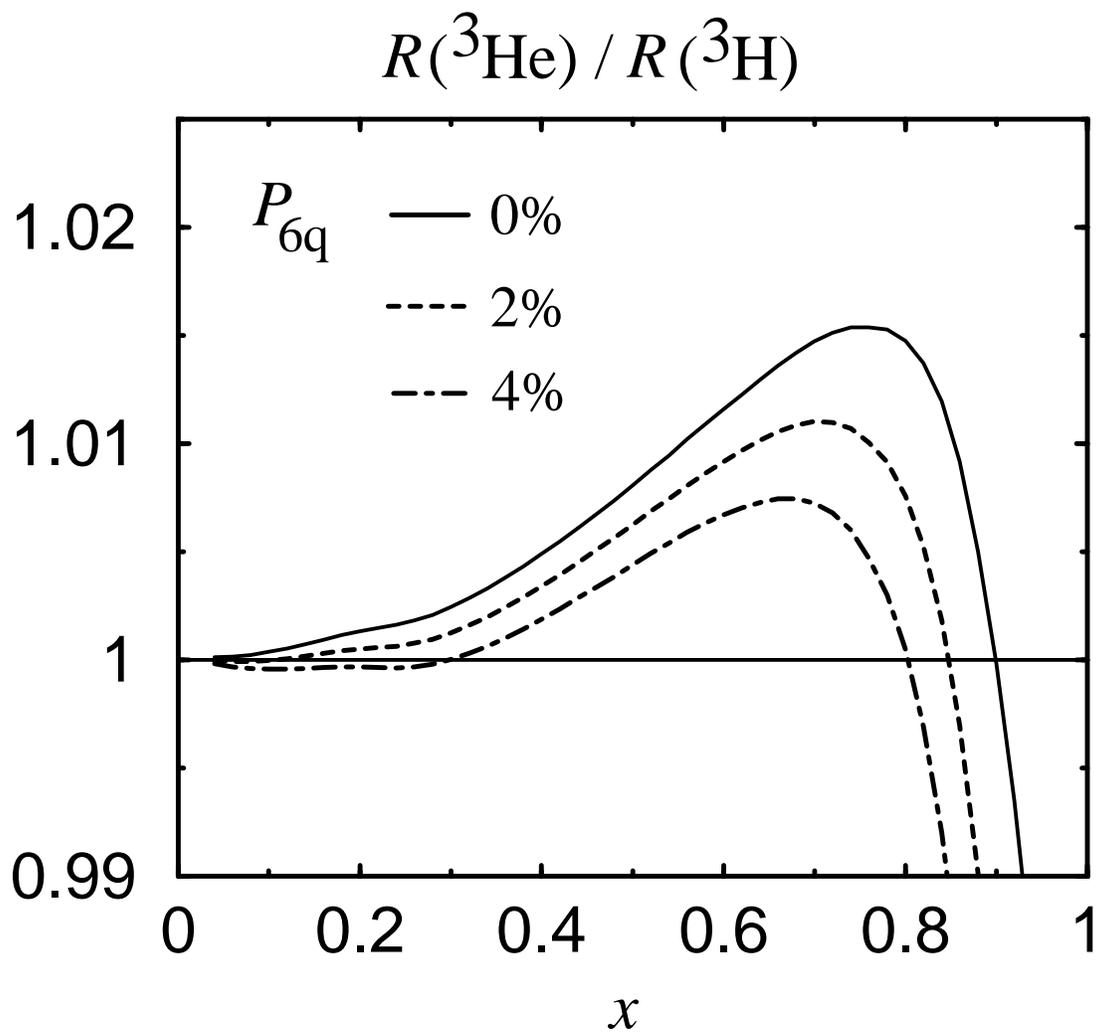,height=14cm}
\vspace*{1cm}
\caption{Ratio of nuclear EMC ratios for $^3$He and $^3$H for the
        Faddeev (PEST) wave function, with $P_{6q} = 0\%$, 2\% and
        4\% six-quark configurations in the $A=3$ wave function.}
\end{center}  
\end{figure}

\begin{figure}  % FIG 12
\begin{center}
\epsfig{figure=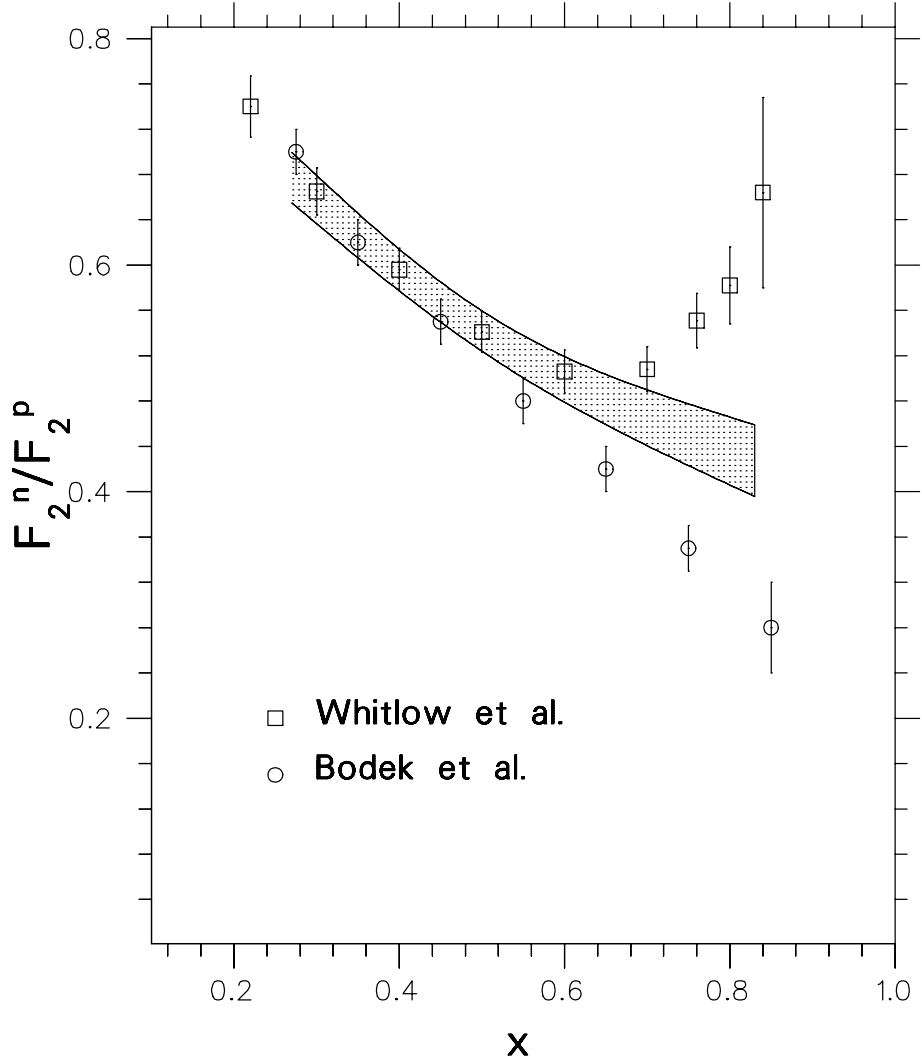,height=14cm}
\vspace*{1cm}
\caption{Two diverging extractions~\cite{WHITLOW,bodek} 
(see text and Fig.~1) of the ratio 
$F_2^n/F_2^p$ from the same SLAC data on inelastic
        proton and deuteron scattering.
        The shaded band represents a $\pm$ one standard deviation
        error band
        for the proposed $^3$H and $^3$He JLab experiment~\cite{phily,pac18}.
        The central values of the band are chosen arbitrarily to 
follow the trend of the analysis of the same data by Melnitchouk and Thomas~\cite{MT}.}
\end{center}
\end{figure}


\begin{thebibliography}{99}
\bibitem{FAD_NOTE}
More accurately, $\varphi_\alpha$ is defined by the equation
$\left\vert\varphi_\alpha \right\rangle
= G_0 V_\alpha \left\vert\Psi \right\rangle$,
where $V_\alpha$ is the potential between particles $\beta$ and
$\gamma$ and $G_0 =(E-H_0)^{-1}$.
\bibitem{VAR_NOTE}
The notation adopted in Sect.~III.B.2 differs slightly from that used
in the Faddeev approach in Sect.~III.B.1, but is consistent with that
used in the literature for variational calculations.
\bibitem{3BODY}
We also omit possible three-body forces since these are expected to
have a negligible effect on ${\cal R}$.
\end{thebibliography}
\end{document}